\documentclass[a4paper,11pt]{article}

\usepackage{epsfig,multicol,amsmath}
\usepackage[T1]{fontenc}
\usepackage{jheppub}
\usepackage{amsmath}     
\usepackage{epsfig,epsf}
\usepackage{graphicx}
\usepackage{rotating}
\usepackage{verbatim}

\def\beq{\begin{equation}}
\def\eeq{\end{equation}}
\def\bea{\begin{eqnarray}}
\def\eea{\end{eqnarray}}

\def\eq#1{{Eq.~(\ref{#1})}}
\def\fig#1{{Fig.~\ref{#1}}}
\newcommand{\bas}{\bar{\alpha}_S}
\newcommand{\as}{\alpha_S}

\newcommand{\Lb}{\left(}
\newcommand{\Rb}{\right)}
\parskip=\itemsep               

\newcommand{\nn}{\nonumber}

\newcommand{\h}{\frac{1}{2}}

\newcommand{\pom}{I\!\!P}

\def\pom{{I\!\!P}}

\title{ CGC/saturation approach for soft interactions at high energy:  a 
two channel model}
\author[a]{ E. ~Gotsman,}
\author[a,b]{ E.~ Levin}
\author[a]{  and U.~ Maor}

\affiliation[a]{Department of Particle Physics, School of Physics and Astronomy,
Raymond and Beverly Sackler
 Faculty of Exact Science, Tel Aviv University, Tel Aviv, 69978, Israel}
\affiliation[b]{Departemento de F\'isica, Universidad T\'ecnica Federico Santa Mar\'ia, and Centro Cient\'ifico-\\
Tecnol\'ogico de Valpara\'iso, Avda. Espana 1680, Casilla 110-V, Valpara\'iso, Chile}
\emailAdd{gotsman@post.tau.ac.il}
\emailAdd{leving@post.tau.ac.il, eugeny.levin@usm.cl}
\emailAdd{maor@post.tau.ac.il}

\abstract{In this paper we continue  the development of a model for strong 
interactions
 at high energy, based on two ingredients:  CGC/saturation approach and 
the
 BFKL Pomeron. In our approach, the unknown mechanism of  confinement
 of quarks and gluons, is characterized by several numerical parameters, 
which are 
  extracted from the experimental data. We demonstrate that the two 
channel model,
   successfully describes the experimental data, including
  both  the value of the elastic slope and the energy  behaviour
  of the single diffraction cross section. We show
 that the disagreement with experimental data of our previous 
single channel eikonal model \cite{GLMNIM},
 stems from the   simplified approach used for the  hadron 
structure,
 and is not related to our principal theoretical input, based on the 
CGC/saturation approach.}

\keywords{ Soft Pomeron, CGC/saturation approach, BFKL Pomeron, diffraction
 at high energies}
\begin{document}
\maketitle
\flushbottom

\pagestyle{empty}

\mbox{}

\pagestyle{plain}

\setcounter{page}{1}


\section{Introduction.}

In this paper we expand our  approach, of analyzing  soft interactions at 
high energy,  
based on two  main ingredients: the Colour Glass
 Condensate(CGC)/saturation effective theory  for high energy QCD (see 
Ref.\cite{KOLEB})
 and references therein), and the BFKL 
Pomeron that describes both soft and hard interactions
 at high energy \cite{BFKL,LI}. The idea that there is  only one BFKL 
Pomeron,
which is  not even a pole in the angular momentum plane (not a Reggeon,
 see Ref.\cite{COL} for details), presumes that the unknown
 mechanism of confinement in QCD of quarks and gluons, is not 
important,
 and that its influence can be mimiced by the determination of several 
parameters
of the CGC/saturation approach, which depend on long distance physics.
 As an example of such a parameter, we mention the behavior of the
 scattering amplitude of the BFKL Pomeron at large impact parameters.
  This  contradicts
  the hope that  confinement would lead to a Pomeron which is
 a Regge pole, (see Ref.\cite{BARTNEW} and references there in). 
 Unfortunately, due to the embryonic state of our understanding of 
confinement in QCD,
 we do not yet have a theoretical tool to differentiate
between 
 these two approaches. Hence, we concentrate our efforts on comparing
the results
 of our approach, with relevant experimental data, hoping that  an 
evaluation
 will allow us to check how viable  our scenario is, and to find the
 specific features where  our approach differs, from  one based on 
  soft Pomeron calculus.
  
Our first attempt \cite{GLMNIM} shows, that we can describe the main
 features of the data, but we found two results in our description
 which imply a potential problem for our approach: the 
result for the elastic 
slope,
  is  much smaller than the experimental measurement at the LHC 
energies, 
and the behaviour  of the cross section for 
single diffraction, which
 displays oscillating saturation as a function of energy.  
  The goal of this paper is to show that these problematic features, 
occur due to our 
 oversimplified model in Ref.\cite{GLMNIM}. In this model,  all shadowing
 corrections stem from two sources: the eikonal rescattering, and 
 the interaction of the BFKL Pomerons taken in the CGC/saturation 
approach.
  In this paper we develop a two channel model instead of the eikonal 
 approximation, which is responsible  for the low mass component in 
diffraction
 production, and which is an essential ingredient of all models of the 
high
 energy hadron scattering,  now on the market 
\cite{DL,GLM1,GLM2,KAP,KMR,OST,NEWGLM,NEWKMR,OSTREC,GLMREV}.

The paper is organized as follows. In the next section we describe
 the main theoretical input  used in this paper.  This
 section reviews the results that have been derived and employed in our
 previous paper \cite{GLMNIM}. In addition, we include a new formula for
  single diffraction dissociation, which is based on the closed form
 solution to  the Balitsky-Kovchegov equation\cite{KOLE}. In section
 3 we discuss
 the structure of our model, and the  phenomenological parameters
 that  have been introduced. In this section, based on theoretical 
considerations,
 we  estimate the 
   range of  values of the parameters.  
 The fourth section is devoted to the description and results
 of the fit.


\section{Theoretical input}

\subsection{Dressed Pomeron}

\begin{figure}[ht]
    \centering
  \leavevmode
      \includegraphics[width=13cm]{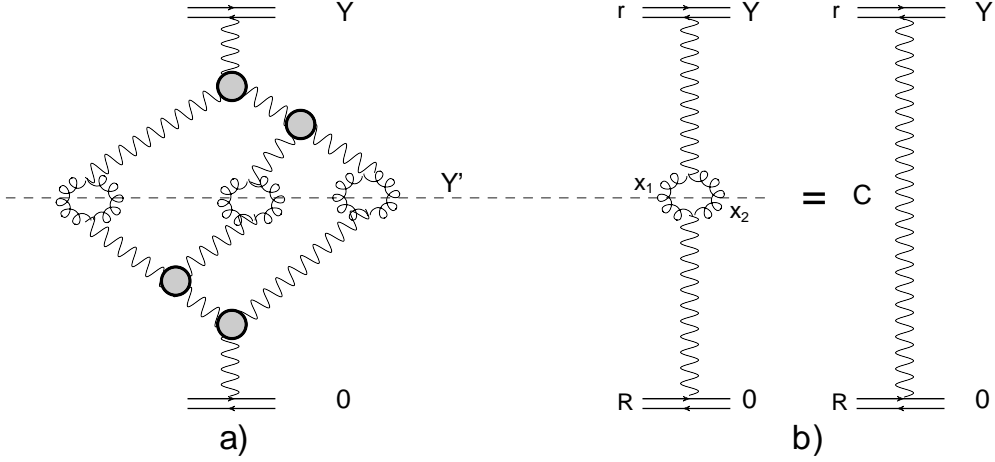}  
      \caption{ MPSI approximation: the simplest diagram(\protect\fig{mpsi}-a)
 and  one Pomeron contribution (\protect\fig{mpsi}-b).
 $C\,\,=\,\,\bas^2/4 \pi$. $ \vec{x}_1 - \vec{x}_2  =
 \vec{r}^{\,'}$. $\h (\vec{x}_1 + \vec{x}_2 ) = \vec{b}^{\,'}$.
  The wavy lines describe  BFKL Pomerons.
 The blobs stand for triple Pomeron vertices.}
\label{mpsi}
   \end{figure}

In the CGC/saturation approach (see Ref.\cite{KOLEB} for the review of 
this 
approach), the scattering amplitude of two dipoles at high energy, is
 described  as the exchange of the dressed Pomeron, which can be 
calculated
 using  the MPSI approximation \cite{MPSI},  displayed in \fig{mpsi}.
  In Ref.\cite{AKLL}   it  was shown, that the MPSI approximation  is 
valid over    a  wide
 range of rapidities:
\beq \label{MPSI2}
Y\,\,\leq\,\,\frac{2}{\Delta_{\mbox{\tiny BFKL}}}\,\ln\Big(\frac{1}
{\Delta^2_{\mbox{\tiny BFKL}}}\Big).
\eeq

 From \fig{mpsi}, it appears that the calculations can be performed in two 
stages: 
first, 
to find the sum of `fan' diagrams for rapidity regions (0, Y') and (Y' Y);
 and , second, to evaluate the diagram of \fig{mpsi}-a  using the 
following
 sum rules for the BFKL Pomeron exchange \cite{GLR,MUSH} (see \fig{mpsi}-b for
 all notations)
\beq \label{POMTUN}
\frac{\as^2}{4 \pi} \,\,G_\pom\Lb Y - 0, r, R; b  \Rb\,=\,\int d^2 r'
  d^2 b' \,G_\pom\Lb Y - Y', r, r', \vec{b} - \vec{b}^{\,'} \Rb \, \,
\,G_\pom\Lb Y' r', R, \vec{b} - \vec{b}^{\,'} \Rb\,\,.
\eeq 

In \eq{MPSI2} $G_\pom $ denotes the Green's function of the BFKL Pomeron.

The first step can be accomplished by  finding a solution to the 
Balitsky-Kovchegov (BK) 
non-linear equation \cite{BK}. The solution has different forms
 in the  three kinematic regions.
\begin{enumerate}
 \item $r^2 Q^2_s\Lb Y, b\Rb \,\ll\,1$, where $Q_s$ denotes the saturation 
scale\cite{GLR,MUQI,MV}. The non-linear corrections are small, and the
 solution is the BFKL Pomeron, which has the following form\cite{BFKL,LI}
 \bea \label{BFKL}
&&G_\pom\Lb Y, r, R; b\Rb\,\,=\,\,\Lb w\,w^* \Rb^\h \sqrt{\frac{\pi}{4\,D\,Y}}
\,e^{ \Delta_{\mbox{\tiny BFKL}}\,Y \,-\,\frac{\ln^2 w\,w^*}{4 \,D\,Y}} 
,\\
&&\mbox{with}~~\Delta_{\mbox{\tiny BFKL}}\,=\,2 \ln 2\,\bas~\mbox{and} ~D\,
=\,14 \zeta(3) \bas \,=\,16.828 \,\bas\nn .
\eea
  $G_\pom\Lb Y, r, R; b\Rb$ denotes the BFKL Pomeron Green
 function,  $\bas$  the QCD coupling,  $r$ and $R$ are  the
 sizes of two interacting dipoles. $Y \,=\,\ln s$, where
 $s = W^2$.  $W$ denotes the energy of the interaction, and $b$  the 
impact 
parameter of the scattering amplitude for two dipoles.  

\beq \label{W}
w\,w^*\,\,=\,\,\frac{r^2\,R^2}{\Lb\vec{b} - \h\Lb\,\vec{r}\,
- \,\vec{R}\Rb\Rb^2
\,\Lb\vec{b} \,+\, \h \Lb\,\vec{r} \,- \,\vec{R}\Rb\Rb^2}.
\eeq
From \eq{BFKL} it is obvious that 
the BFKL Pomeron is not  a pole in   angular 
momentum, but a branch cut, since its Y-dependence  
has an additional $\ln s$ term; it also does not reproduce the exponential
 decrease at large $b$, which follows from the general properties of
 analyticity and unitarity,  for  the exchange of the BFKL 
Pomeron \cite{FROI}.

  \item $r^2 Q^2_s\Lb Y, b\Rb \,\sim\,1$ (vicinity of the saturation scale).
 The scattering amplitude has the following form\cite{MUT,IIM}
  \beq \label{TI1}
   A\,\,\equiv\,\,G_\pom\Lb  z\Rb\,\,=\,\,\mbox{ Const}\Lb r^2 \,
 Q^2_s\Lb Y, b\Rb\Rb^{1 - \gamma_{cr}},
   \eeq 
 where $\mbox{ Const}$ denotes a constant, and  where the critical 
anomalous 
dimension $\gamma_{cr}$, can be found from
 \beq \label{GACR}
\frac{\chi\Lb \gamma_{cr}\Rb}{1 - \gamma_{cr}}\,\,=\,\, -
 \frac{d \chi\Lb \gamma_{cr}\Rb}{d \gamma_{cr}}~~~\,\,\,\mbox{and}\,\,\,
~~~\chi\Lb \gamma\Rb\,=\,\,2\,\psi\Lb 1 \Rb\,-\,\psi\Lb \gamma\Rb\,-\,\psi\Lb
 1 - \gamma\Rb.
\eeq
 \item $r^2 Q^2_s\Lb Y, b\Rb \,>\,1$ ( inside the saturation domain).
 The breakthrough that allows us to develop  phenomenology based on
 the CGC/saturation approach,  was published in \cite{LEPP}. In this
 paper a simple    approximation to the numerical solution of
 the BK equation was  found, which is of  the 
form
 
 \beq \label{SOLNU}
N^{BK}\Lb G_\pom\Lb z\Rb\Rb \,\,=\,\,a\,\Lb 1
 - \exp\Lb -  G_\pom\Lb z\Rb\Rb\Rb\,\,+\,\,\Lb 1 - a\Rb
\frac{ G_\pom\Lb z\Rb}{1\,+\, G_\pom\Lb z\Rb},
\eeq 
 with $a = 0.65$.
      \end{enumerate}
    
Using the solution of \eq{SOLNU} we  calculated   the diagrams shown in 
\fig{mpsi}-a using \eq{POMTUN} (see Refs.\cite{GLMNIM,LEPP}). The
 result of this calculation gives the following expression for the Green
 function of the dressed Pomeron:
 
\bea\label{DP1}
&&G^{\mbox{dressed}}_\pom\Lb Y- Y_0, r,  R, b \Rb\,\,=\,\, \\
&&a^2\Bigg\{ 1 \,-\,\exp\Lb - T\Lb Y- Y_0, r,  R, b \Rb \Rb \,\Bigg\}\,+\,2 a (1 -  a) \frac{ T\Lb Y- Y_0, r,  R, b \Rb }{
1\,\,+\,\,T\Lb Y- Y_0, r,  R, b \Rb }\nn\\
  && +\,\,( 1 -  a)^2 \,\Bigg\{1 - \exp\Lb \frac{1}{ T\Lb Y- Y_0, r,  R, b \Rb }\Rb\,\frac{1}{T\Lb Y- Y_0, r,  R, b \Rb }\,
  \Gamma\Lb 0,   \frac{1}{
T\Lb Y- Y_0, r,  R, b \Rb }\Rb   \Bigg\}\nn ,
 \eea 
 where $\Gamma\Lb  x \Rb$ is the incomplete Euler
 gamma function (see {\bf 8.35} of Ref.\cite{RY}). The
 function $T\Lb Y- Y_0, r,  R, b \Rb$ can be found
 from \eq{POMTUN}, and has the form:
 \beq \label{DP2}
 T\Lb Y- Y_0, r,  R, b \Rb \,\,=\,\,\frac{\bas^2}{4 \pi}\,G_\pom\Lb
 z \to 0 \Rb \,\,=\,\,\phi_0 \Lb r^2 Q^2_s\Lb R, Y,b \Rb\Rb^{1 -
 \gamma_{cr}}\,=\,\phi_0 S\Lb b \Rb e^{\lambda (1 - \gamma_{cr}) Y},
 \eeq
 where we used two inputs: $r = R$ and $Q^2_s\,=\,\Lb 1/(m^2R^2)\Rb\,S\Lb m\,
, b \Rb
 \, \exp\Lb \lambda \,Y\Rb$.  The parameter  $\lambda$, in leading order 
of 
perturbative QCD, is given by $\lambda\,=\,\bas
 \chi\Lb \gamma_{cr}\Rb/(1 -
 \gamma_{cr})$ .  The parameter $m$ and the function $S\Lb m \, ,b\Rb$
  originate from  non-perturbative QCD contributions,  and are 
parmeterized as:
 \beq \label{SDI}
S \Lb m \,b\Rb \,\,=\,\,\frac{m^2}{\pi^2} \,e^{- m\,b}~~~~\mbox{where}~~~~\int
 d^2 b \,S\Lb b\Rb\,=\,1
\eeq
   $\phi_0$ can be calculated from the initial conditions using \eq{BFKL}. 
  Since, we do not know these conditions,  we will consider $\phi_0$ as 
an 
additional phenomenological parameter.
 \subsection{Single diffraction}
 The equation for the single diffraction production was  proposed
 more than a decade ago \cite{KOLE}. In Ref.\cite{KOLE} it was shown
 that the equation has the same form as the BK equation\cite{BK,JIMWLK}
 for the function
 \beq \label{SD1}
 G\Lb Y, Y_0,r, b \Rb\,\,=\,\,2\,N\Lb Y, r, b \Rb \,\,-\,\,N^{SD}\Lb Y, Y_0,
 r, b\Rb ,
 \eeq
 where 
 \beq \label{SD2}
 \sigma_{diff}\Lb Y,Y_0,r\Rb\,\,=\,\,\int d^2 b \,N^{SD}\Lb Y, Y_0, r, b\Rb
 \eeq
 is the cross section for diffraction production, with the rapidity gap 
($Y_{gap} \,=\,Y \,-\,Y_M$) larger than $Y_0$( $Y_{gap} \,\geq\,Y_0$).
 
\begin{figure}[ht]
    \centering
  \leavevmode
      \includegraphics[width=7cm]{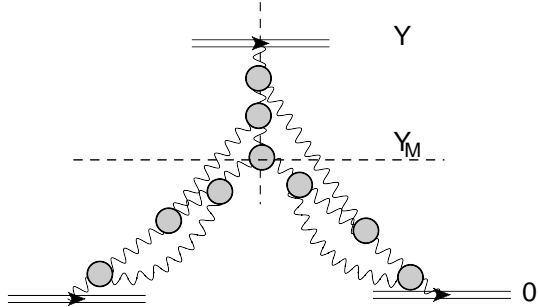}  
      \caption{ MPSI approximation: the simplest diagrams for 
 single diffraction production.
  The wavy lines describe  BFKL Pomerons.
 The blobs stand for triple Pomeron vertices. The dotted line denotes
 the cut Pomeron. $Y_M\,=\,\ln\Lb M^2/s_0\Rb$, where $M$ is the mass of
 produced particles and $s_0 $ is the scale taken to be of the order
 of $1\,GeV^{2}$.}
\label{sd}
   \end{figure}

 $ N\Lb Y, r, b \Rb$ is the imaginary part of the elastic amplitude,
 other notations are   clarified in \fig{sd}.
 
 The difference between the  elastic amplitude and $ G\Lb Y, Y_0,r, b 
\Rb$, is 
only in the initial condition, which for $  G\Lb Y, Y_0,r, b \Rb$ is given 
by the following equation:
 \beq \label{ICSD}
   G\Lb Y, Y_0 \,= \,Y, r,  b \Rb  \,\,=\,\, 2\,N\Lb Y, r, b \Rb \,\,-\,\,N^2\Lb Y, r, b \Rb    
   \eeq
   where the last term denotes the elastic cross section.
  
   Bearing in mind that the solution to \eq{ICSD}  is given by 
\eq{SOLNU}, 
we can obtain 
a solution for  single diffractive production. The
 cross section for the production of a bunch of hadrons with a mass
 from $M_{min}$ to $M_{max}$
 can be written as
   \beq \label{SD2}
   \sigma_{diff}\Lb Y,Y_{max},Y_{min},r\Rb\,\,=\,\,\int d^2 b \,\tilde{N}^{SD}\Lb Y, Y_{max}, Y_{min},r;  b \Rb   \eeq
   where  the amplitude $,\tilde{N}^{SD}\Lb Y, Y_{max}, Y_{min} r;  b \Rb $ takes the form:
\beq \label{ND}
\tilde{N}_{SD}\Lb Y, Y_{max}, Y_{min}, r ; b\Rb\,\,=\,\,
N^{BK}\Lb TT\Lb  Y, Y_{max}, b\Rb\Rb\,\,-\,\,N^{BK}\Lb TT\Lb  Y,Y_{min},b\Rb\Rb\eeq
where $Y_{max} = \ln\Lb M^2_{max}/s_0\Rb$ and $Y_{min} = \ln\Lb M^2_{min}/s_0\Rb$.

For $ TT\Lb  Y, Y_M, b\Rb$ we have the following expression
\beq \label{TT}
TT\Lb  Y,Y_M,b\Rb\,\,=\,\,\Bigg( 2 \,G^{\mbox{\tiny dressed}}\Lb
 T\Lb Y - Y_M, b\Rb\Rb
\,\,-\,\,\Big(G^{\mbox{\tiny dressed}}\Lb T\Lb Y -  Y_M,
 b\Rb\Rb\Big)^2\Bigg)\,e^{ \Lb 1 - \gamma_{cr}\Rb\,\lambda\,\Lb Y_M\Rb} .
\eeq

 \subsection{Double diffraction}
 We use  $s$-channel unitarity, to obtain the expression for the
  cross section for double diffractive production.
 Indeed, the unitarity constraints  for the dressed Pomeron given
 by the diagrams of \fig{mpsi}-a take the form

\beq \label{NDD1}
2 \,G^{\mbox{\tiny dressed}}\Lb T\Lb Y, b \Rb\Rb\,\,=\,\,
G^{\mbox{\tiny dressed}}\Lb 2\, T\Lb Y; b\Rb\Rb\,\,+\,\,N_{DD}\Lb Y; b 
\Rb\,
\eeq
 where $G^{\mbox{\tiny dressed}}\Lb 2\, T\Lb Y; b\Rb\Rb$ describes
 all inelastic processes that are generated by the dressed Pomeron
 exchange. One can check this formula for the exchange of a single
 BFKL Pomeron. The general proof that 
 the inelastic cross section due to the exchange of the dressed Pomeron,
 is given by  $G^{\mbox{\tiny dressed}}\Lb 2\, T\Lb Y; b\Rb\Rb$ can be
 found in Refs. \cite{BKKMR,LEPR}.
 
 From \eq{NDD1} the cross section of  double  diffractive
 production is equal to
 
 \beq \label{NDD2}
 \sigma_{dd}\,\,=\,\,\int d^2 b\,\Big\{ 2 G^{\mbox{\tiny dressed}}\Lb
 T\Lb Y, b \Rb\Rb\,\,-\,\,
G^{\mbox{\tiny dressed}}\Lb 2\, T\Lb Y; b\Rb\Rb \Big\} .
\eeq

 In \eq{NDD2} we integrated over all possible
 masses without any restriction.  We do not expect  the cross
 section for  double diffractive production of small masses to be large, 
and we 
believe
 that  most of the contribution in this region of masses stem from the 
Good
 Walker mechanism \cite{GW}, which we consider in the next  section.

 \section{ Main formulae and its phenomenological parameters}
 


\subsection{Two channel approximation}
 In the previous section we  reviewed  the theoretical
 input from  the CGC/saturation approach, used for calculating  the
 Green function of the resulting Pomeron. In this section we
 discuss a model approach making two simplifications: one, that
 the eikonal formula for the hadron scattering amplitude  satisfies
  $s$-channel unitarity; and two, that the simplified two channel model 
 describes diffractive production in the  low mass region. In
 this model, we replace the rich structure of the  diffractively produced
 states, by a single  state, with  wave function $\psi_D$.
  The observed physical 
hadronic and diffractive states can then be written  
\beq \label{MF1}
\psi_h\,=\,\alpha\,\Psi_1+\beta\,\Psi_2\,;\,\,\,\,\,\,\,\,\,\,
\psi_D\,=\,-\beta\,\Psi_1+\alpha \,\Psi_2;~~~~~~~~~
\mbox{where}~~~~~~~ \alpha^2+\beta^2\,=\,1 .
\eeq 

Functions $\psi_1$ and $\psi_2$  form a  
complete set of orthogonal
functions $\{ \psi_i \}$ which diagonalize the
interaction matrix ${\bf T}$
\beq \label{GT1}
A^{i'k'}_{i,k}=<\psi_i\,\psi_k|\mathbf{T}|\psi_{i'}\,\psi_{k'}>=
A_{i,k}\,\delta_{i,i'}\,\delta_{k,k'}.
\eeq
The unitarity constraints can be written as
\beq \label{UNIT}
2\,\mbox{Im}\,A_{i,k}\left(s,b\right)=|A_{i,k}\left(s,b\right)|^2
+G^{in}_{i,k}(s,b),
\eeq
where $G^{in}_{i,k}$ denote the contributions of all non 
diffractive inelastic processes,
i.e. it is the summed probability for these final states to be
produced in the scattering of a state $i$ off a state $k$. In \eq{UNIT} 
$\sqrt{s}=W$, is the energy of the colliding hadrons, and $b$ denotes the 
impact  parameter.
A simple solution to \eq{UNIT} at high energies, has the eikonal form 
with an arbitrary opacity $\Omega_{ik}$, where the real 
part of the amplitude is much smaller than the imaginary part.
\beq \label{GT3}
A_{i,k}(s,b)=i \Lb 1 -\exp\Lb - \frac{\Omega_{i,k}(s,b)}{2}\Rb\Rb ;
\eeq
\beq \label{GT4}
G^{in}_{i,k}(s,b)=1-\exp\Lb - \Omega_{i,k}(s,b)\Rb.
\eeq
\eq{GT4} implies, that the probability that the initial projectiles
$(i,k)$ will reach the final state interaction unchanged, regardless, of 
the initial state re-scatterings, is  given by
$P^S_{i,k}=\exp \Lb - \Omega_{i,k}(s,b) \Rb$.
\par

The physical observables in this model can be written as follows
\bea
\mbox{elastic~~ amplitude}:&~~&a_{el}(s, )\,=\,i \Lb \alpha^4 A_{1,1}\,+\,2 \alpha^2\,\beta^2\,A_{1,2}\,
+\,\beta^4 A_{2,2}\Rb; \label{OBS1}\\
\mbox{elastic observables}:&~~~&\sigma_{tot}\,=\,2 \int d^2 b\, a_{el}\Lb s, b\Rb;~~~\sigma_{el}\,=\, \int d^2 b \,|a_{el}\Lb s, b\Rb|^2;\nn\\
\mbox{optical~~~\, theorem}:&~~~&2 \,\mbox{Im} A_{i,k}(s,t=0)\,= \,2 \int d^2 b\, \mbox{Im} A_{i,k}(s,b)\,
=\,\sigma_{el} + \sigma_{in} \,=\,\sigma_{tot};\label{OBS11}\\
\mbox{single ~ diffraction}:&     & \sigma^{GW}_{sd}\,=\,\int\,d^2 b \, \Lb\alpha \beta\{-\alpha^2A_{1,1}
+(\alpha^2-\beta^2)A_{1,2}+\beta^2 A_{2,2}\}\,\Rb^2;\label{OBS2}\\
\mbox{double \,\,diffraction}:&     &\sigma^{G W}_{dd}\,\,=\,\,\int d^2 b
 \,\, \alpha^4\beta^4
\left\{A_{1,1}\,-\,2\,A_{1,2}\,
+\, A_{2,2} \right\}^2.\label{OBS3}
\eea

We denote by `GW' the Good -Walker component, that is responsible for 
 diffraction in the  small mass region.

In the eikonal approach we parametrize the arbitrary functions
 $\Omega_{i k}\Lb s, b \Rb$  in  the form
\beq \label{OMEGA}
\Omega_{i k}\Lb s, b \Rb\,\,=\,\,\int d^2 b'  d^2 b'' g_i\Lb
 m_i, b'\Rb\,g_k\Lb m_k b''\Rb\, G^{\mbox{\tiny dressed}}\Lb
 T\Lb Y, \vec{b} - \vec{b}^{\,'}  - \vec{b}^{\,''} \Rb\Rb ,
\eeq
where the vertex $g_i\Lb m_i, b'\Rb$ is parameterized as
\beq \label{G}
g_i\Lb m_i, b'\Rb\,\,=\,\,g_i \,S_\pom \Lb m_i, b'\Rb , 
~~~~~~\mbox{where}~~~~S_\pom \Lb m_i, b'\Rb \,\,=\,\, 
\,\frac{1}{4\pi}\,m^3_i\,b'\,K_1\Lb m_i\,b' \Rb .
\eeq
$S_\pom\Lb b \Rb$ in \eq{G} is the Fourier transform of the form factor
 $1/(1 + q^2/m^2_i)^2$  and $K_1\Lb z \Rb$ denotes the modified Bessel 
function 
of the second kind (the McDonald function, see  formulae {\bf 8.4} in 
Ref.\cite{RY}).

\subsection{Small phenomenological parameters and net diagrams}
\begin{figure}[ht]
    \centering
  \leavevmode
      \includegraphics[width=7cm]{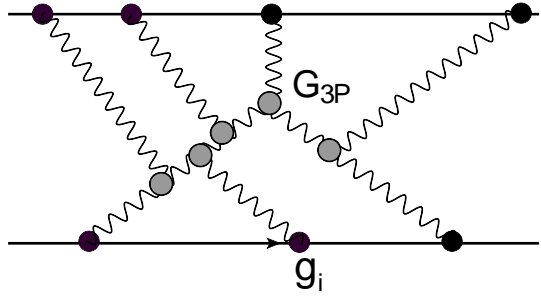}  
      \caption{ The net diagram that gives the largest contribution
 if $g_i\,\gg\,G_{3 \pom}$.
  The wavy lines describe  BFKL Pomerons.
 The blobs denote the triple Pomeron vertices. }
\label{net}
   \end{figure}

The fits to experimental data (see Refs.\cite{GLMREV,GLMNIM} and
 references therein) led to an unexpected result viz: the value of 
vertices for
 Pomeron-hadron interactions ($g_i$), 
extracted from the fit of the data, turn out to be much larger
 than the vertex of triple Pomeron interaction ( $G_{3\pom} $  see \fig{net}).
   
Due to this small parameter $G_{3\pom}/g_i \,\ll\,1$, we can improve the 
eikonal 
approximation, and sum a more general class of diagrams, the
 so called net diagrams shown in \fig{net}. The analytical
 expression for the sum of the `net' diagrams has been found
 (see Ref. \cite{GLMREV}).  Using the inequality observed in our 
previous paper\cite{GLMNIM}:
 viz.  $m$ in \eq{SDI} turns out to be much larger than both $m_1$ and 
$m_2$. 
  Assuming this inequality to hold,  we can simplify the integrals over 
$b'$ and $b''$ in 
\eq{OMEGA}, re-writing this equation in the form
\bea \label{OMEGASIM}
\Omega_{i k}\Lb s, b \Rb\,\,&=&\,\,\int d^2 b'  d^2 b'' g_i\Lb m_i,
 b'\Rb\,g_k\Lb m_k b''\Rb\, G^{\mbox{\tiny dressed}}\Lb T\Lb Y,
 \vec{b} - \vec{b}^{\,'}  - \vec{b}^{\,''} \Rb\Rb\,\nn\\
&=&\,
\Lb \int d^2 b''\,G^{\mbox{\tiny dressed}}\Lb T\Lb Y, b'' \Rb\Rb\Rb\,\int d^2
 b' g_i\Lb m_i, b'\Rb\,g_k\Lb m_k, \vec{b} - \vec{b}^{\,'}\Rb .
\eea

The summation of the `net' diagram, is then given by the following 
simplified
 expression (see Ref.\cite{GLMREV} for details)
\bea
\Omega\Lb Y; b\Rb~~&=& ~~ \int d^2 b'\,
\,\,\,\frac{ g_i\Lb\vec {b}'\Rb\,g_k\Lb\vec{b} -
 \vec{b}'\Rb\,\widetilde{G}^{\mbox{\tiny dressed}}\Lb T\Rb
}
{1\,+\,1.29\,\widetilde{G}^{\mbox{\tiny dressed}}\Lb T\Rb\left[
g_i\Lb\vec{b}'\Rb + g_k\Lb\vec{b} - \vec{b}'\Rb\right]} , \label{OM}\\
\eea
where
\beq \label{GT}
\widetilde{G}^{\mbox{\tiny dressed}}\Lb T\Rb\,\,=\,\,\int \,d^2 b
 \,G^{\mbox{\tiny dressed}}\Lb T\Lb Y, b \Rb\Rb.
\eeq

The coefficient 1.29 results from the extraction of the value of 
$G_{3\pom}$ 
from the CGC/saturation approach, that has been considered in the previous 
section.

\subsection{Diffraction production in the region of large mass}

In this section we also  include, in the process  of diffraction 
production,
 the mechanism of production that originates from the dressed Pomeron, and 
has been discussed in section 2.

For single diffraction the  large mass contribution can be written 
\bea\label{SDLM}
&&\sigma^{\mbox{large mass}}_{sd}\,\,=\,\,\,\,2 
\int d^2 b \,\\
&&\left\{\,\alpha^6\,A^{sd}_{1;1,1}\,
e^{- 2\,\Omega^D_{1,1}\Lb Y;b\Rb}\,\,+\,\alpha^2\beta^4 A^{sd}_{1;2,2}\,
e^{- 2\,\Omega^D_{1,2}\Lb Y;b\Rb} + 2\,\alpha^4\,\beta^2 \,A^{sd}_{1;1,2}\,
e^{- \Lb \Omega^D_{1,1}\Lb Y;b\Rb
+ \Omega^D_{1,2}\Lb Y; b \Rb\Rb} \right. \nn\\
&&\left.\,\,+\,\,\beta^2\,\alpha^4 \,A^{sd}_{2;1,1}\,
e^{- 2\,\Omega^D_{1,2}\Lb Y;b\Rb}\,\,+\,\,2\,\beta^4\alpha^2
\,A^{sd}_{2;1,2}\,e^{- \Lb \Omega^D_{1,2}\Lb Y;b\Rb
+ \Omega^D_{2,2}\Lb Y; b \Rb\Rb}\,\,+\,\,\beta^6\,\,
A^{sd}_{2; 2,2}\,e^{- 2\,\Omega_{2,2}\Lb Y;b\Rb} \right\}. \nonumber
\eea
where 
\beq \label{OMD}
\Omega^D_{i,k}\Lb Y;b\Rb\,\,=\,\, \int d^2 b'\,
\,\,\,\frac{ g_i\Lb\vec {b}'\Rb\,g_k\Lb\vec{b} - \vec{b}'\Rb\,\bar{G}^{\mbox{\tiny dressed}}\Lb T\Rb
}
{\Big(1\,+\,1.29\,\bar{G}^{\mbox{\tiny dressed}}\Lb T\Rb\left[
g_i\Lb\vec{b}'\Rb + g_k\Lb\vec{b} - \vec{b}'\Rb\right]\Big)^2}
\eeq
\bea \label{AD}
&&A^{sd}_{i; k,l}\Lb Y,Y_{max},Y_{min}; b\Rb\,\,= \\
&&\,\, \int d^2 b'\,  \sigma_{diff}\Lb Y,Y_{max},Y_{min},1/m\Rb d^2 
\,b'\,g_i g_k g_l 
S_\pom\Lb b', m_i\Rb\, S_\pom\Lb \vec{b} - \vec{b}', m_k\Rb \,S_\pom\Lb
 \vec{b} - \vec{b}', m_l\Rb ,\nn
\eea
where  $   \sigma_{diff}\Lb Y,Y_{max},Y_{min},r\Rb $ is given by \eq{SD2}, $ Y = \ln \Lb s/s_0\Rb$ , $Y_{max} = \ln\Lb M^2_{max}/s_0\Rb$ and $Y_{min} = \ln\Lb M^2_{min}/s_0\Rb$. $M_{max}$ and $M_{min}$ are the largest and smallest mass produced in the diffractive processes.

\eq{SDLM} has simple physical meaning: each term is the product of
 probability to produce a mass diffractively from the dressed Pomeron
 (term $\exp\Lb- \sum \Omega\Rb$), and the probability of the process
 of single diffraction, from the dressed Pomeron ($A_{i;k, l}$).

For the double diffraction production at large mass  we have
\beq \label{DDLM}
\sigma^{\mbox{\tiny large mass}}_{dd}\,\,=\,\,\int\,d^2 b\,
\left\{\alpha^4\,A^{dd}_{1,1}\,e^{- 2 \Omega^D_{1,1}
\Lb Y;b\Rb}\,+ \,2 \alpha^2\,\beta^2 A^{dd}_{1,2}\,
e^{- 2 \Omega^D_{1,2}\Lb Y;b\Rb}\,+\,\,\beta^4\,A^{dd}_{2,2}\,
e^{- 2 \Omega^D_{2,2}\Lb Y;b\Rb}\,\right\}.
\eeq
\beq \label{ADD}
A^{dd}_{i,k}\,\,=\,\,\int d^2\, b\, g_i\,g_k \,S^{i,k}_{DD}
\Lb b \Rb \sigma_{dd}\Lb Y\Rb~~~\mbox{where}~~S^{i,k}_{DD}
\Lb b \Rb\,=\,\int d^2 b'\,S_p\Lb b', m_i \Rb\,S_p\Lb \vec{b} - \vec{b}',
 m_k\Rb ,
\eeq
where $\sigma_{dd}$ is given by \eq{NDD2}.

\subsection{Estimates for the values of the phenomenological parameters}
We have  two sets of phenomenological parameters,  which
 need to be  determined by fitting to the experimental data. 
The first set is related to the description of the dressed Pomeron:
 $\phi_0, \lambda$  and $m$. The parameter $\phi_0$, in principle, could 
be found
 from the solution of the BFKL equation at low energy. Unfortunately, 
we do not know
 the initial condition for  BFKL evolution. This is the reason that we  
extract this
 parameter by  fitting to the experimental data.  The value of
$\phi_0$ should  be
   of the order of $\as$, and therefore, we expect that this
 parameter will be small. $\lambda$ determines the energy dependence
 of the saturation scale. We know the theoretical value of
 $\lambda\,=\,2\bas \Big(\psi\Lb 1\Rb - \psi \Lb  \h\Rb\Big)/(1 -
 \gamma_{cr})$ where $\psi(z)$ denotes the Euler psi-function (see 
formulae
  8.36 of Ref.\cite{RY}) and $\gamma_{cr} = 0.37$ (see
 Ref.\cite{KOLEB}). On the other hand, the value of $\lambda
 \,\approx\,0.3$ has been extracted from DIS and nucleus-nucleus
 scattering for the energy dependence of the saturation
 scale\cite{LR,KLN,AAQ}. The mass $m$ plays a twofold role:
 it determines the impact parameter dependence, and it gives the
 size of the typical dipole in a hadron. This is a dimensional
 parameter, whose value is determined by  non-perturbative
 QCD, and, at the moment, we have no theoretical input for this quantity. 
In 
all
 our formulae, we use the intuitive assumption that the mass $m$, is the
 largest  mass in our model.

The second set of parameters: $g_i$ and $m_i$ as well as values of
 $\alpha$,   are associated with the description of the wave functions in 
our
 two channel model. This set is of non-perturbative origin, and has to
 be determined by  fitting to the data. In our formulae we assume
 that $g_i \,\gg\,G_{3\pom}$ and $m \,\gg\, m_i$. This assumption is
 based on our past experience with  soft Pomeron models  for high 
energy
 scattering \cite{GLMREV}.


\section{The result of the fit}

\subsection{Cross sections}

   We have eight parameters to be determined by fitting to the 
experimental 
data  on total, inelastic and elastic cross sections,  single and double
 diffractive production cross sections, and  the slope of the forward 
elastic
 differential cross section. The value of the  minimal  energy for data 
that we use is
  $W =$ 0.546 TeV,  as starting from this energy  the 
CGC/saturation
 approach, is able to describe the data on inclusive production
 in proton-proton collisions (see Ref.\cite{LR}) . For lower energies,
 saturation occurs in ion-ion and proton-ion collisions, but
 not in proton-proton collisions \cite{KLN}.


\begin{table}
\begin{tabular}{|l|l|l|l|l|l|l|l|l|}
\hline
model &$\lambda $ & $\phi_0$ &$g_1$ ($GeV^{-1}$)&$g_2$ ($GeV^{-1}$)& $m(GeV)$ &$m_1(GeV)$& $m_2(GeV)$ & $\beta$\\
\hline
2 channel  & 0.38& 0.0019 & 110.2&  11.2 & 5.25&0.92& 1.9 & 0.58 \\
\hline
1 channel &0.323&0.019 & 25.7& n/a&6.35& 0.813&  n/a &n/a \\
\hline
\end{tabular}
\caption{Fitted parameters of the model.}
\label{t1}
\end{table}

\begin{figure}[h]
\centering
\begin{tabular}{c c c}
\includegraphics[width=0.4\textwidth, height=5.5cm]{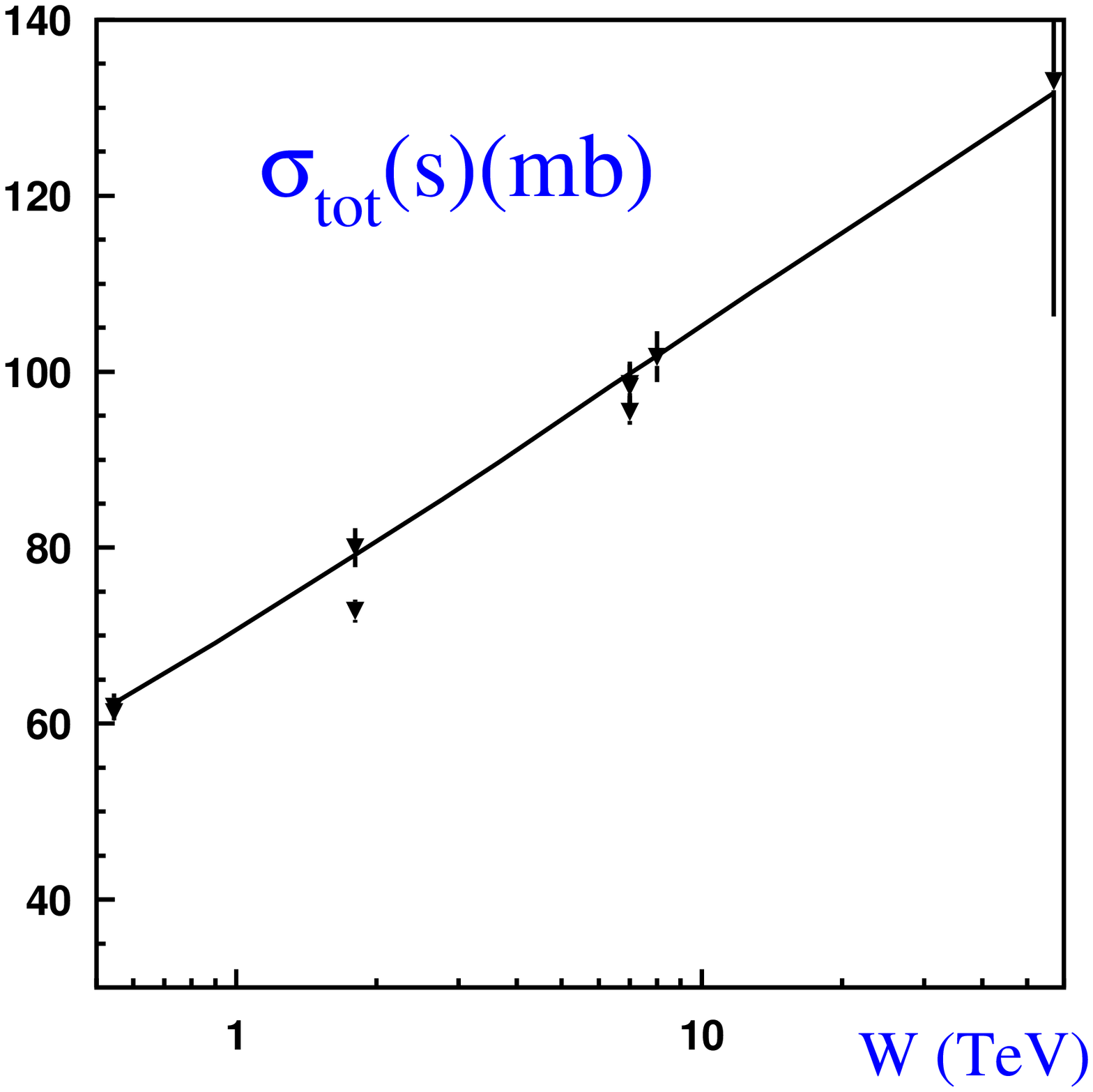}&~~~~~~~~~&
\includegraphics[width=0.4\textwidth,height=5.5cm]{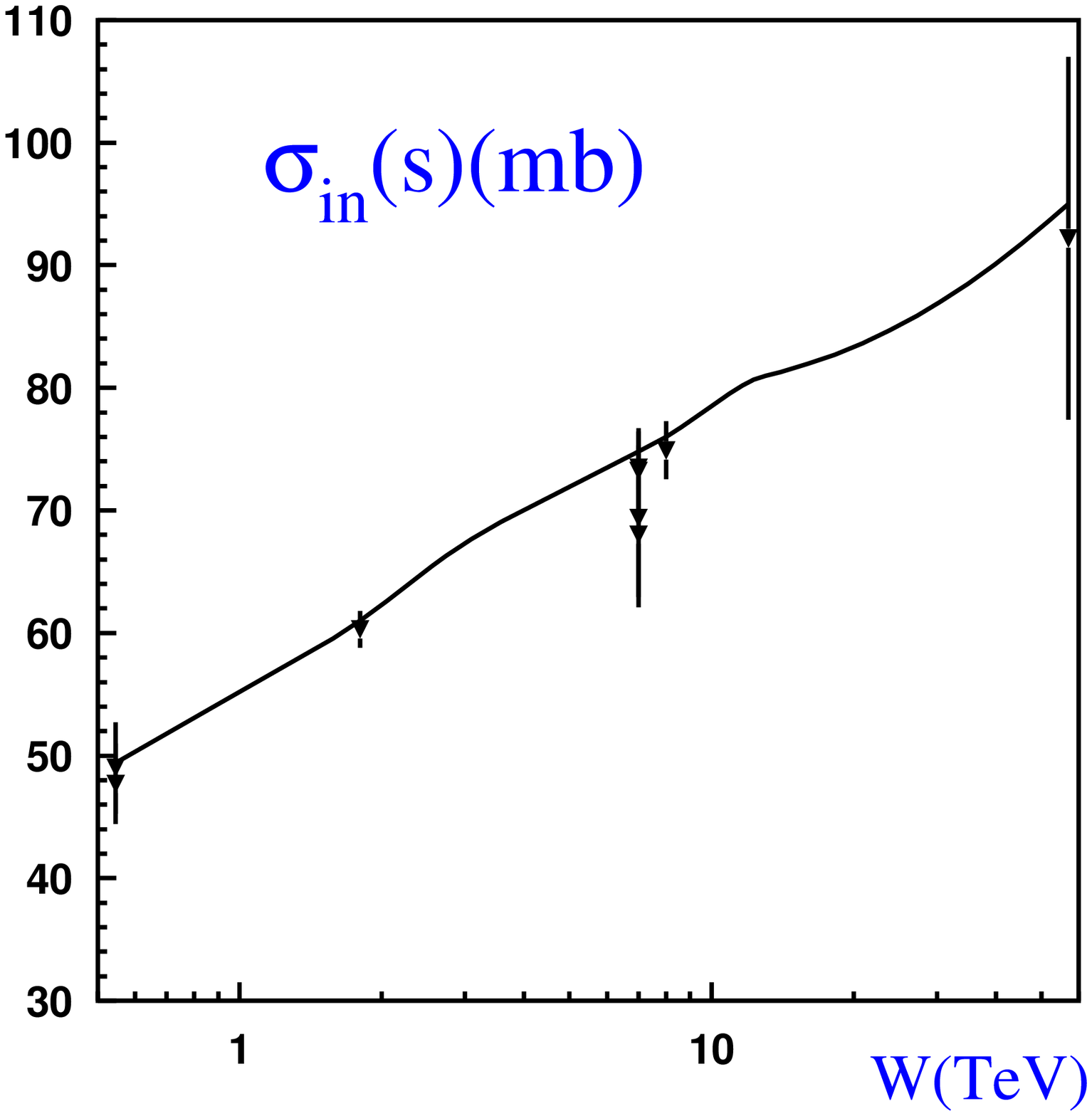}\\
\fig{fit}-a &  &\fig{fit}-b \\
\includegraphics[width=0.4\textwidth,height=5.5cm]{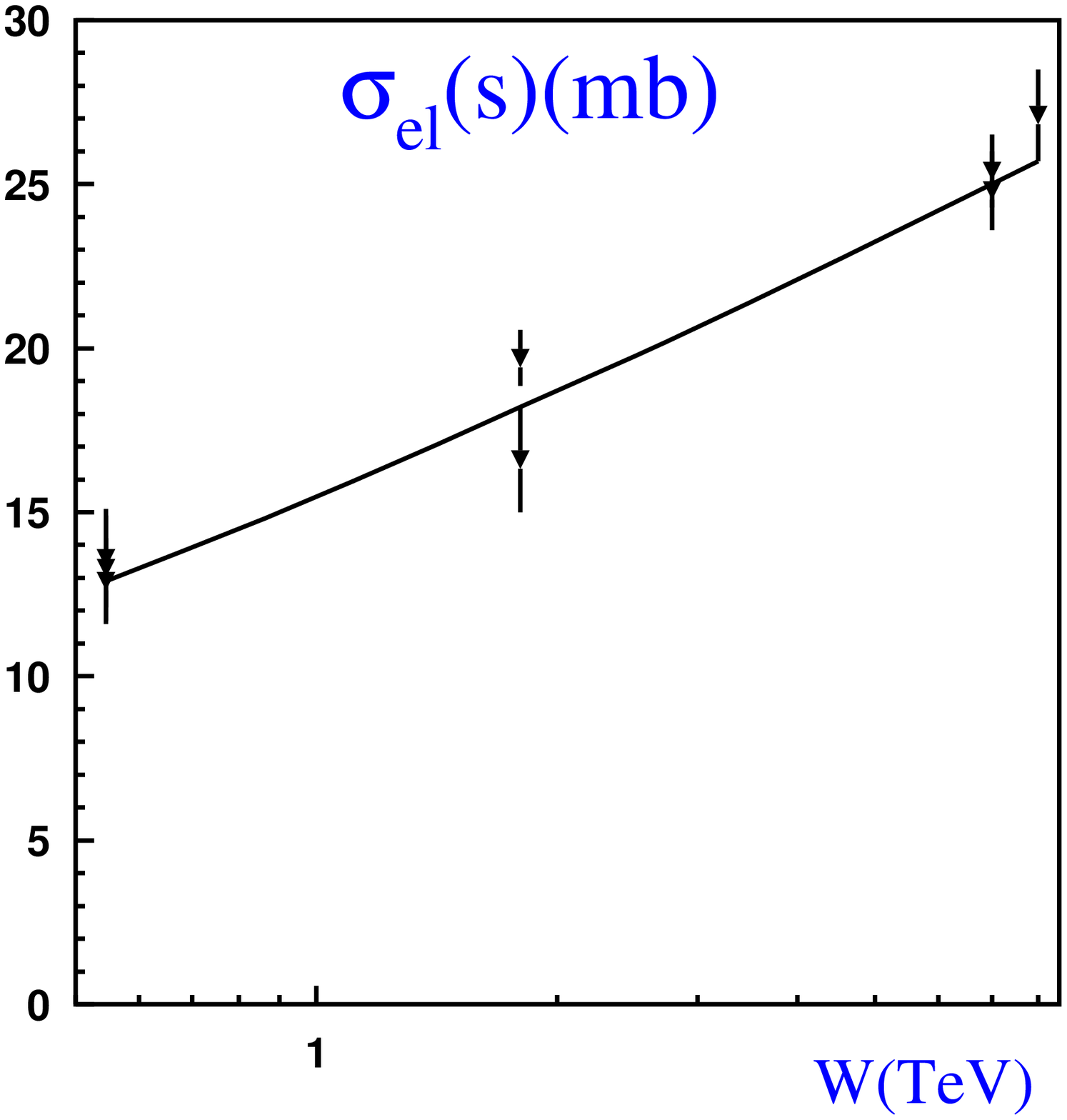}&  &
 \includegraphics[width=0.4\textwidth,height=5.5cm]{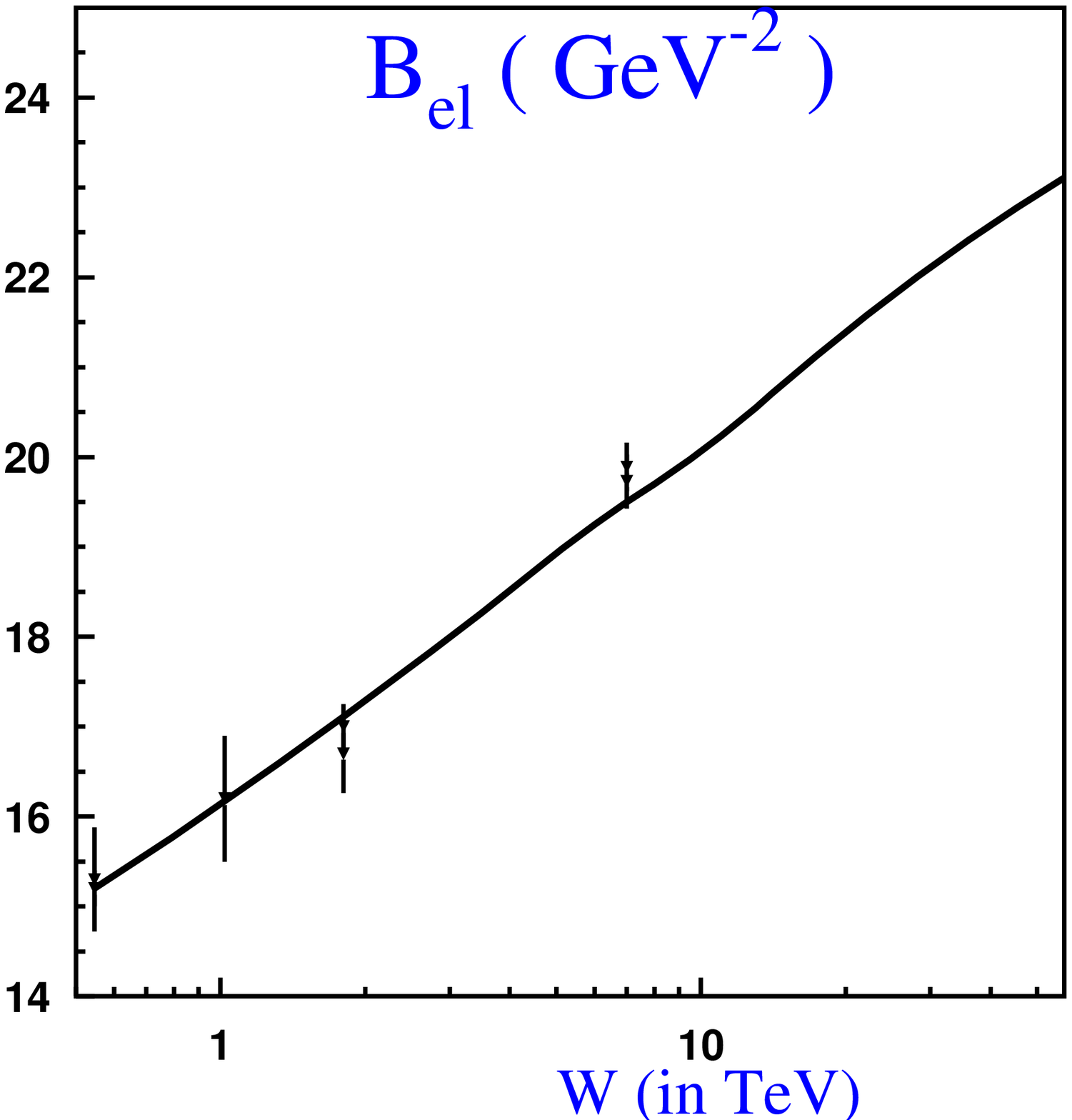}\\
 \fig{fit}-c & &\fig{fit}-d\\
 \includegraphics[width=0.4\textwidth,height=5.5cm]{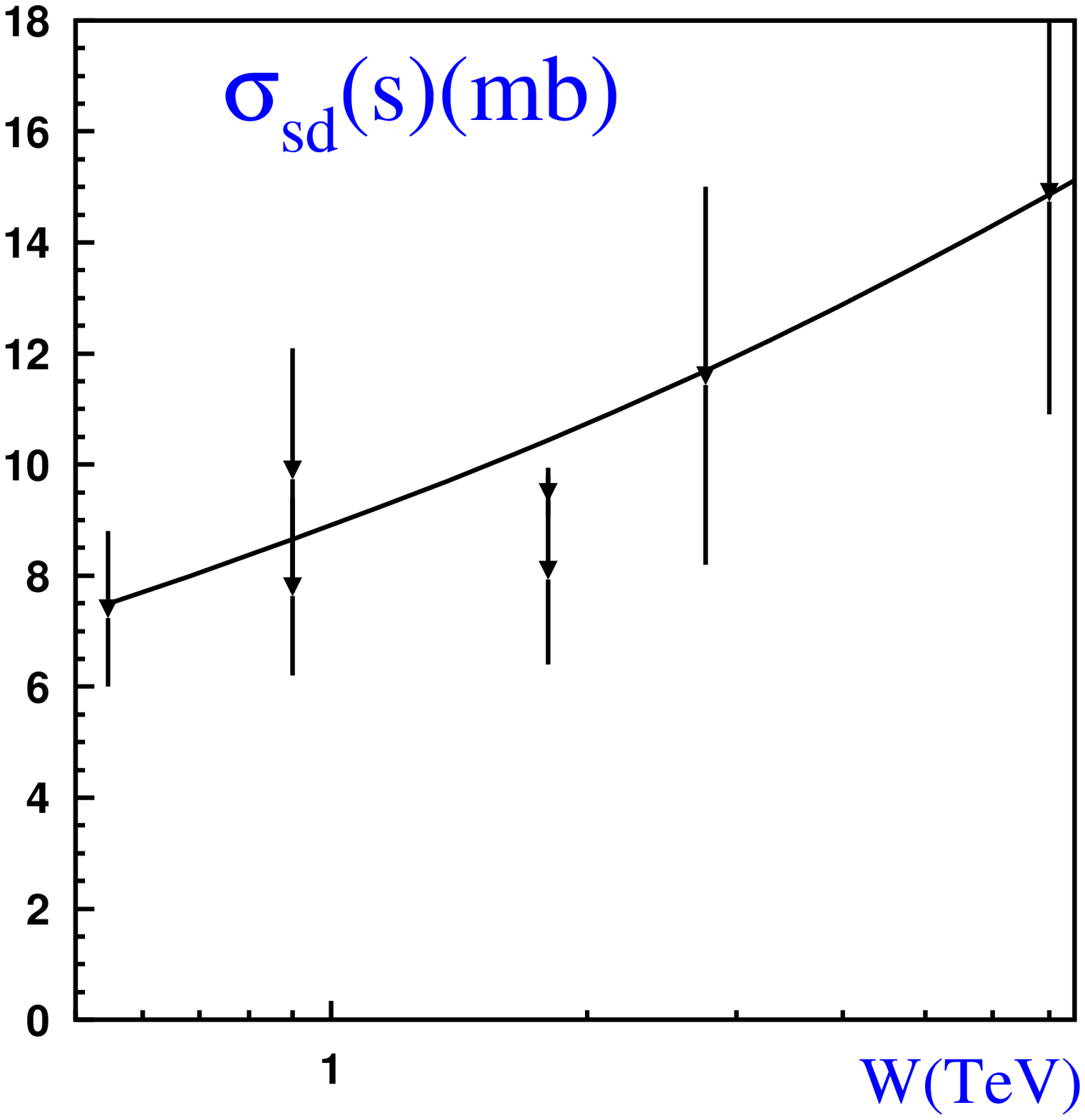}& &
\includegraphics[width=0.4\textwidth,height=5.5cm]{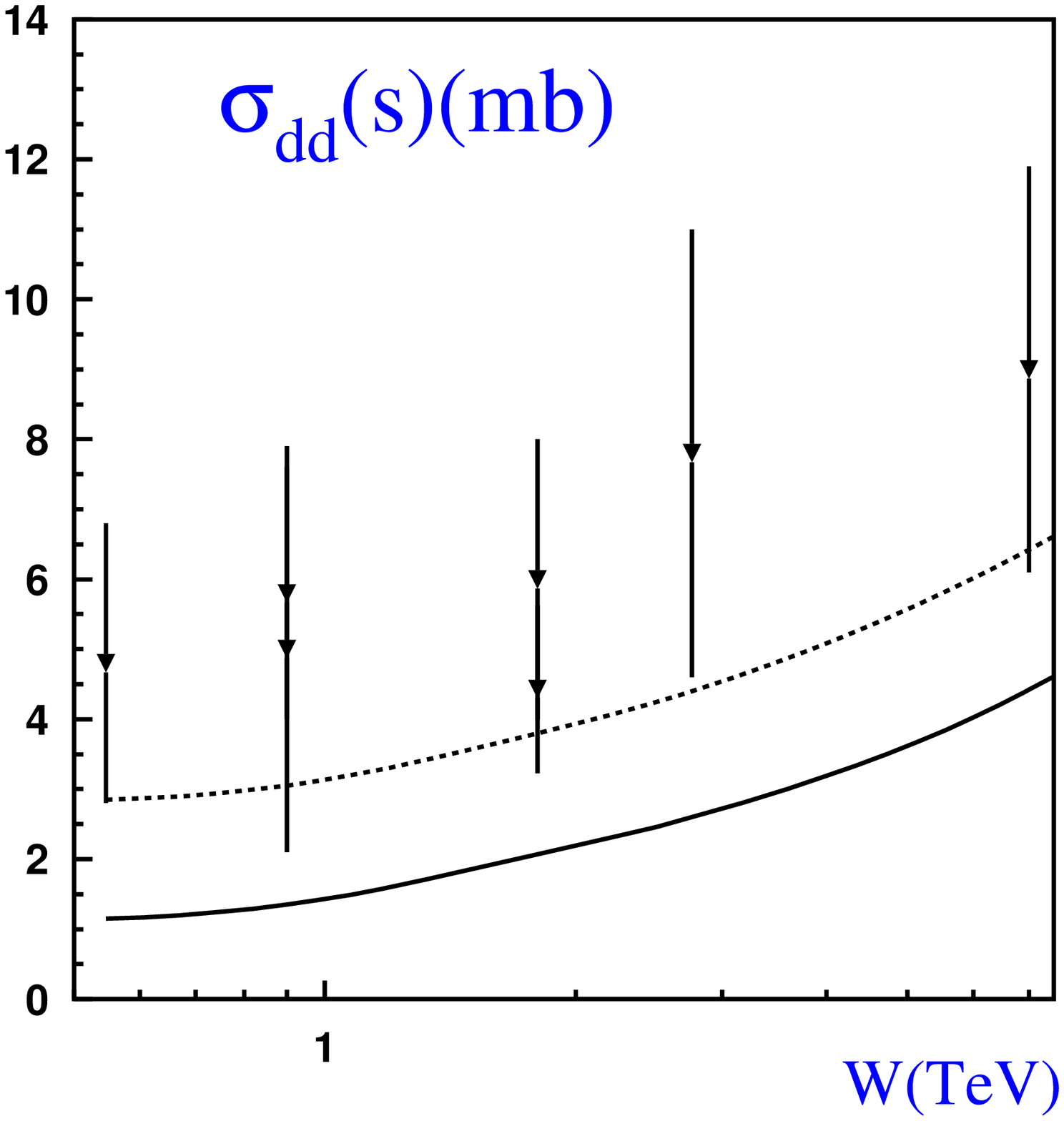}\\
\fig{fit}-e& &\fig{fit}-f \\
\end{tabular}
\caption{Comparison with the experimental data: the energy behaviour
 of the total (\protect\fig{fit}-a), inelastic  (\protect\fig{fit}-b),
 elastic cross sections   (\protect\fig{fit}-c), as well as the elastic slope ($B_{el}$,\protect\fig{fit}-d) and 
   single diffraction 
(\protect\fig{fit}-e) and 
double diffraction (\protect\fig{fit}-f) cross sections.
The solid lines show our present fit. The data has been taken from            
Ref.\protect\cite{PDG}
for energies less than the LHC energy. At the LHC energy for total and
elastic cross
section we use TOTEM data\protect\cite{TOTEM} and for single and double
 diffraction
cross sections are taken from Ref.\protect\cite{ALICE}.
The dotted line in (\protect\fig{fit}-f) is based on Eq.(4.1), (see text 
following this equation).
}
\label{fit}
\end{figure}

The quality of the fit  can be judged from \fig{fit}.
  The fitted parameters are tabulated in Table I.

From Table 1 we see that the qualitative features of the two channel model 
are similar to those of the one channel model
  \cite{GLMNIM}. The values of all the  parameters have
 the same 
hierarchy i.e. $g_i \,\,\gg\,\,G_{3\pom}$
 ( $G_{3 \pom} =
 1.29\, GeV^{-1}$ in our approach) and $m \,\gg\,m_i$. We recall that
 this hierarchy is necessary for our formulae to be valid. $\lambda 
\approx
 0.3 $,  follows from the estimates of the energy dependence of the
 saturation momentum from DIS and ion-ion scattering experiments. $\phi_0$
 is small, as is expected from QCD.

We obtain a good description of the data, including the energy behaviour
 of the elastic slope and the single diffraction cross section.  Therefore,
 we have elucidated the problems that occurred in the one channel 
model\cite{GLMNIM}.
   We show in this paper, that the
 oscillatory behavior of the single diffraction, as well as the weak
 energy dependence of the elastic slope were  artifacts of the
 oversimplified  one channel model,  and are not associated with our main 
theoretical
 input: the  CGC/saturation approach.

A deficiency in our two channel model is the description of the double 
diffractive cross section, as can be seen in  \fig{fit}. This failure
is due to the  small Good-Walker contribution for the production 
of low
 masses. It predicts a cross section  of about  1 mb.  To understand
 the sources of this small value, we note that \eq{OMEGA} leads to the 
cross
 section for  double diffraction, that can be calculated using the
 factorization relation (see Ref.\cite{NEWKMR} for example).
\beq \label{FR}
\sigma^{GW}_{dd} \,=\,\frac{\Lb \sigma^{GW}_{sd}\,B_{sd}\Rb^2}{ \Lb \sigma_{el} \,B_{el}\Rb B_{dd}}
\eeq
Using our parameters in Table I we found that $B_{sd}~ \approx B_{el}~
 \approx ~B_{dd}$, where $B$ are  the slopes of the various cross 
sections. 
 Based on our estimates for elastic and single diffraction cross section,
 we obtain $\sigma_{dd}^{GW} \approx 3 \,mb$ at W = 7 TeV.  
 If we evaluate the cross section of Good-Walker component using \eq{FR}
 we obtain reasonable description of the experimental data (see dotted
 curve in \fig{fit}-f). We wish to stress that the shadowing
 corrections that follows from the CGC/saturation approach, do not
 violate the factorization properties of \eq{FR}. The source of the 
violation of
 \eq{FR}, is the net diagrams, which in our model incorporate
  the interactions of the dressed Pomerons. To illustrate
 the influence of this interaction, we calculated the slopes at W = 7 TeV.
 We obtain : $B_{el} = 19.45 \,GeV^{-2}; B_{sd} = 25.08 
 \,GeV^{-2}$ and $B_{dd}\,=\,44\,GeV^{-2}$.  Using \eq{FR} with
 these values, decreases the estimated $\sigma_{dd}$ by a factor of almost 
three .
 Hence, the main effect of the dressed Pomerons interactions is to
  increase the typical $b$, in the impact parameter distribution for
 double diffractive cross sections.

On the other hand, the small cross section for double diffraction arises
 naturally  in other  theoretical   approaches \cite{NEWKMR},
 and the cross section, measured by the TOTEM collaboration\cite{TOTEMDD}
  in the restricted region of  the produced mass is rather
 small ($\sigma_{dd}\,=\, 116 \pm 25\,\mu b$ for $\Delta Y_M
 = 1.8 $ and $ | Y - Y_{max} | =  4.7$). Bearing this in mind,
 we are not too  disappointed with our description of
 the double diffractive cross section.

In Table 2 we present  the results of our model for the various cross
 sections at different energies. Note,  that for single and double 
diffraction,
 the large mass region is responsible for substantial contributions (more 
than
 half of the cross section).  These
 contributions originate from the structure of the dressed Pomeron, and
 are the main theoretical input from the CGC/saturation approach.

\begin{center}
\begin{table}
\begin{tabular}{|l|l|l|l| l l |l l|}
\hline
W &$\sigma_{tot} $& $\sigma_{el}$(mb) &$B_{el}$&~~~single& diffraction~~ &~~~~double& diffraction~~~ \\
(TeV)   &  (mb)  &        (mb)              &          $(GeV^{-2})$& $\sigma^{LM}_{sd}$ (mb)  &$\sigma^{HM}_{sd}$ (mb)& $\sigma^{LM}_{dd}$ (mb)&$\sigma^{HM}_{dd}$ (mb)\\
\hline
0.576 &62.3& 12.9&15.2 & 5.64& 1.85& 0.7&0.46\\
\hline
0.9 & 69.2 &15&16 &6.25& 2.39& 0.77&0.67 \\
\hline
1.8&79.2&18.2&17.1&7.1&3.35 & 0.89&1.17 \\
\hline
2.74 &85.5&20.2&17.8&7.6&4.07 &0.97&1.62\\
\hline
7 &99.8&25&19.5&8.7& 6.2&1.15&3.27\\
\hline
8 & 101.8&25.7&19.7&8.82&6.55 &1.17&3.63\\
\hline
13 & 109.3&28.3&20.6&9.36& 8.08 & 1.27&5.11\\
\hline
14 & 110.5&28.7&20.7&9.44& 8.34 & 1.27 &5.4\\
\hline
57 & 131.7&36.2 &23.1&10.85&15.02 & 1.56 &13.7\\
\hline
\end{tabular}
\caption{ The values of cross sections versus
 energy. $\sigma^{LM}_{sd}$  and $\sigma^{LM}_{dd}$
 denote the cross sections for  diffraction dissociation
 in the low mass region, for single and double diffraction, which stem
 from the Good-Walker mechanism. While  $\sigma^{HM}_{sd}$  and 
$\sigma^{LM}_{dd}$
 are used for diffraction in high mass, coming from the dressed Pomeron
 contributions.}

\label{t2}
\end{table}
\end{center}

\subsection{Amplitudes}

It is instructive to compare the two models which are both  based on 
CGC/saturation 
approach: the one channel model developed in Ref.\cite{GLMNIM} and 
the two 
channel 
model of this paper.  In \fig{amp} we plot  the amplitudes $A_{ik}$ for 
the two 
channel model and elastic amplitude for the one channel model, at 
different energies,
 as functions of  the impact parameter, b.
 From \fig{amp}-a  we note that one of our amplitude ($A_{2,2}$ is 
rather small at 
7\,TeV, while the amplitude $A_{11}$  reaches the unitarity limit for 
$b \,\leq\,1.5 \,fm$. It is interesting to compare this behaviour  with 
the results of
 two channel model,  based on the Pomeron as a Regge pole
 \cite{GLMREV}. One  notes a drastic difference in the dependence
 on the impact parameter, as well as the relative  values of the
 amplitude.  Qualitatively, the Pomeron interaction
 leads to  stronger shadowing corrections, than the CGC/saturation 
approach.

\begin{figure}[ht]
    \centering
  \leavevmode
      \includegraphics[width=7cm]{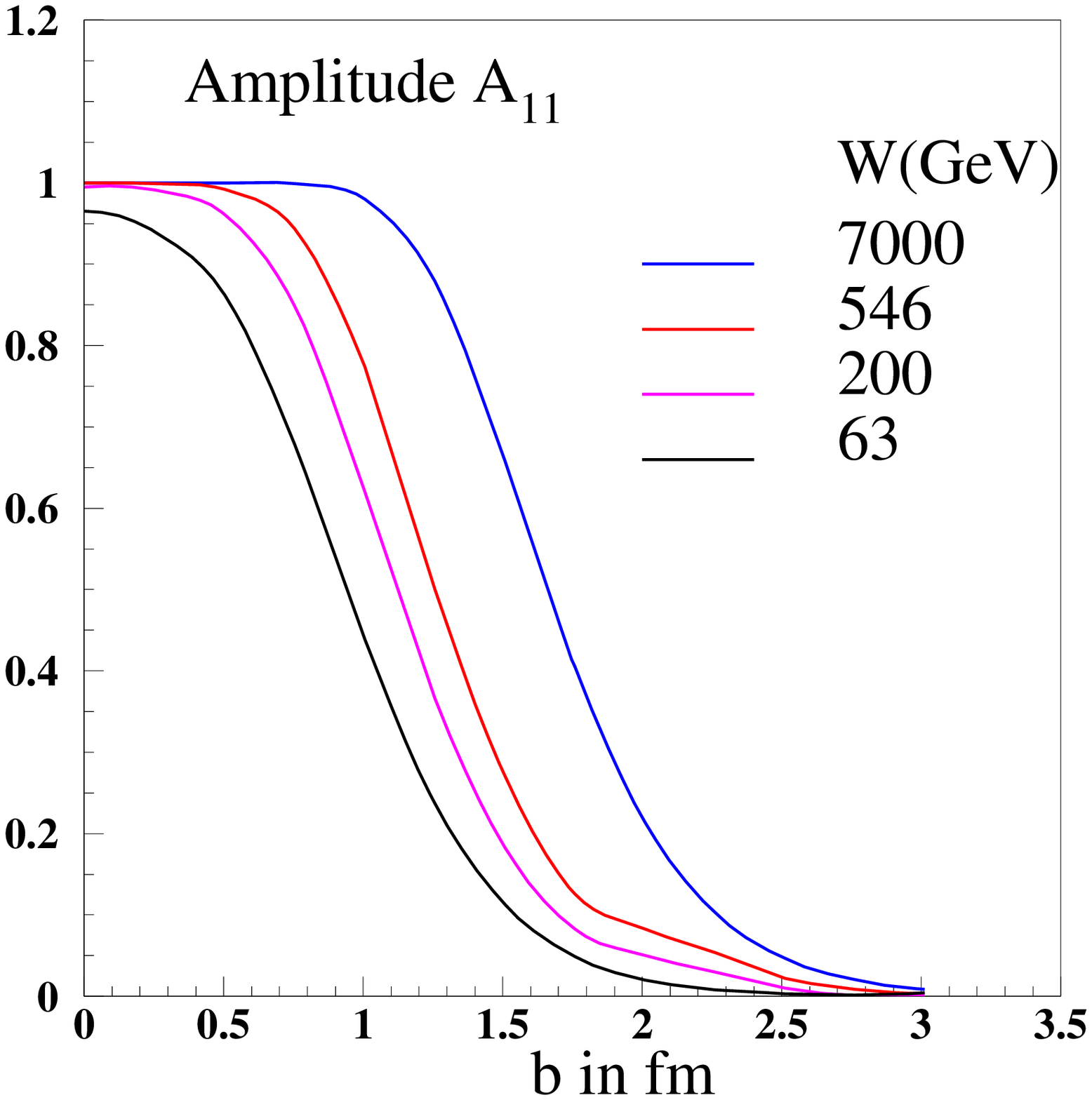}  
      \caption{The  impact parameter dependence of the `black' $A_{1,1}$ 
amplitude, as function of energy W.  }
\label{a11w}
   \end{figure}

\begin{figure}[h]
\centering
\begin{tabular}{c c c }
\includegraphics[width=0.4\textwidth,height=5.5cm]{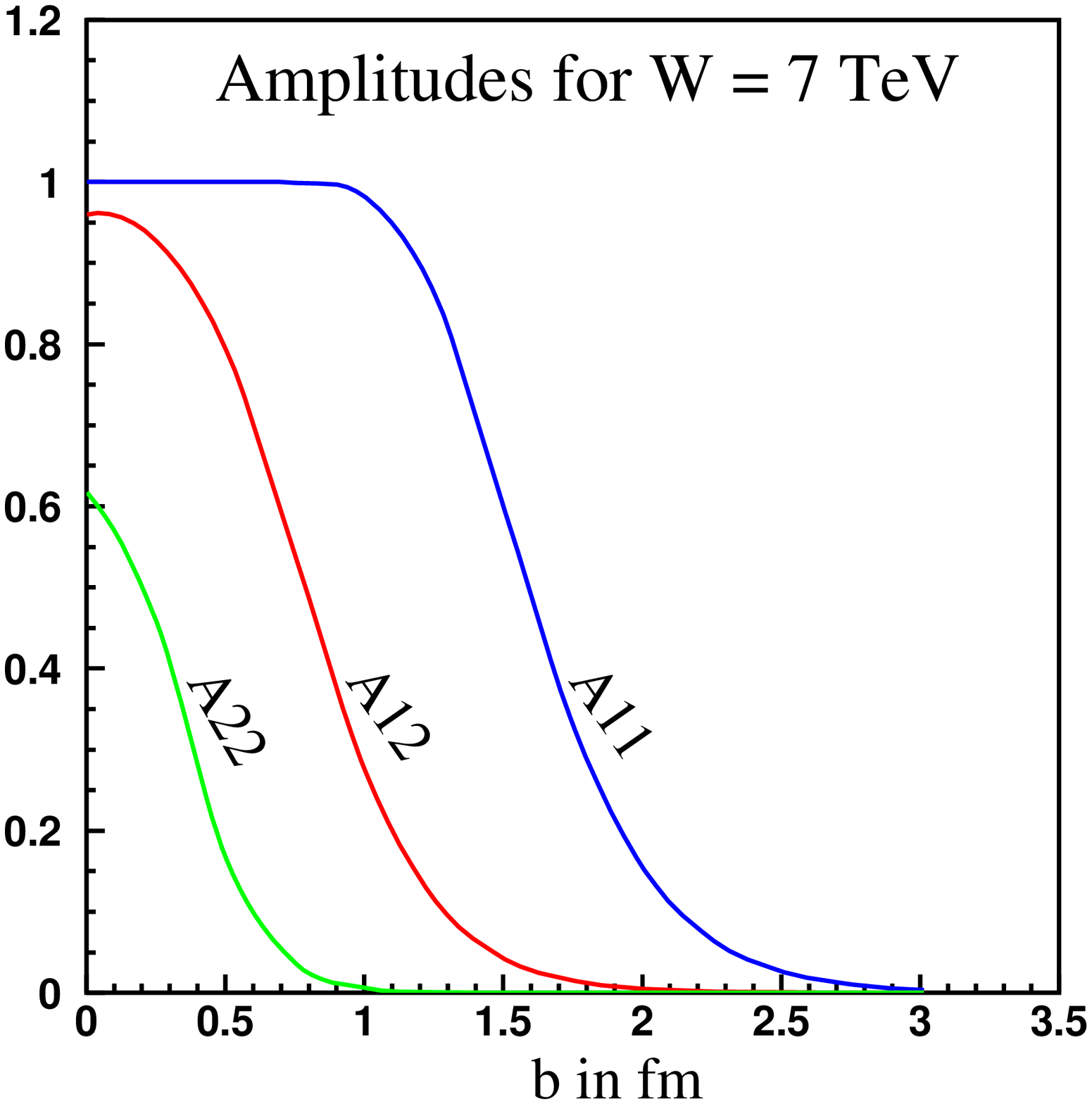}&~~~&\includegraphics[width=0.4\textwidth,height=5.5cm]{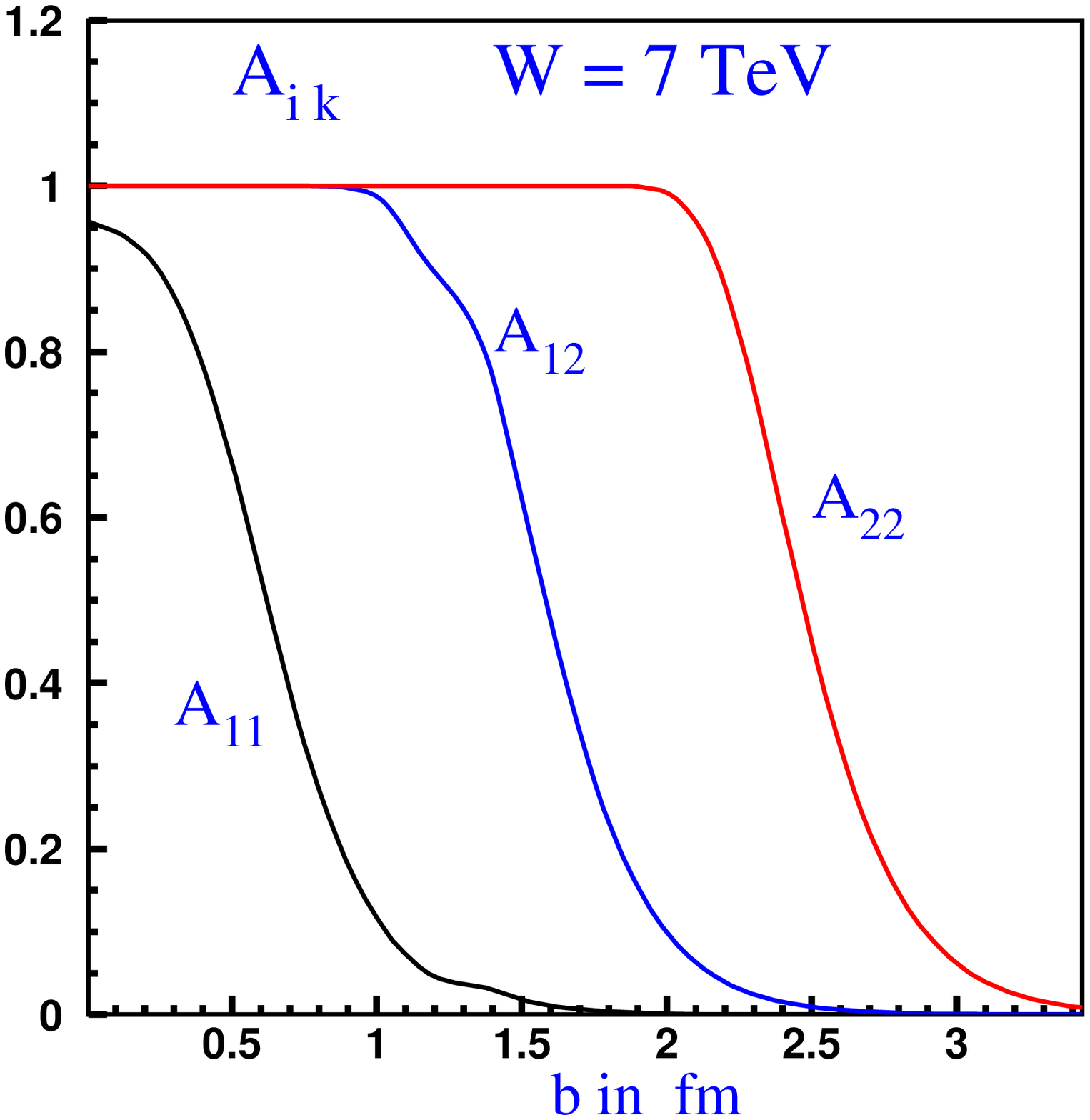}\\
\fig{amp}-a &   &\fig{amp}-b\\
\includegraphics[width=0.4\textwidth,height=5.6cm]{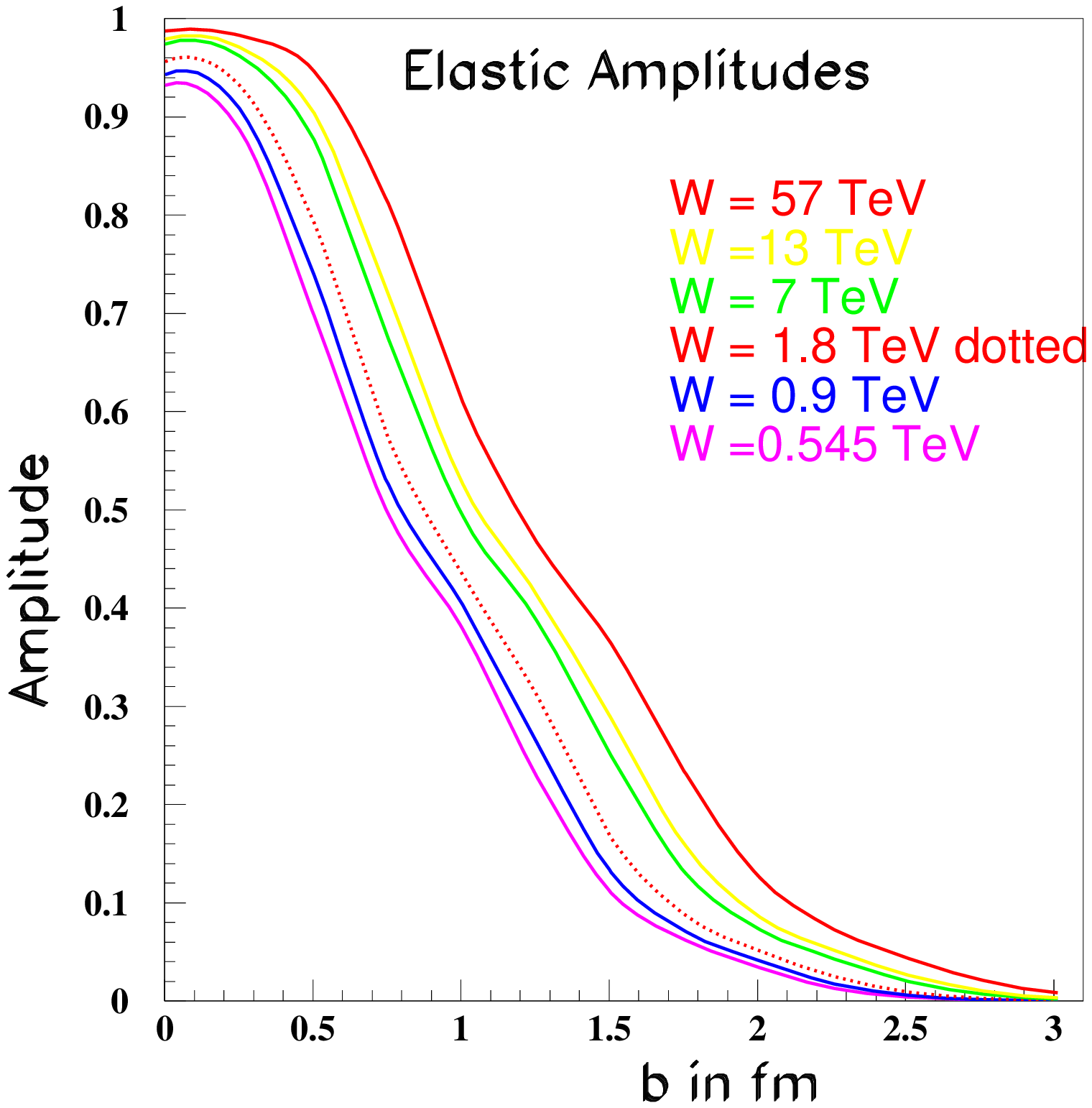}&  &
 \includegraphics[width=0.4\textwidth,height=5.5cm]{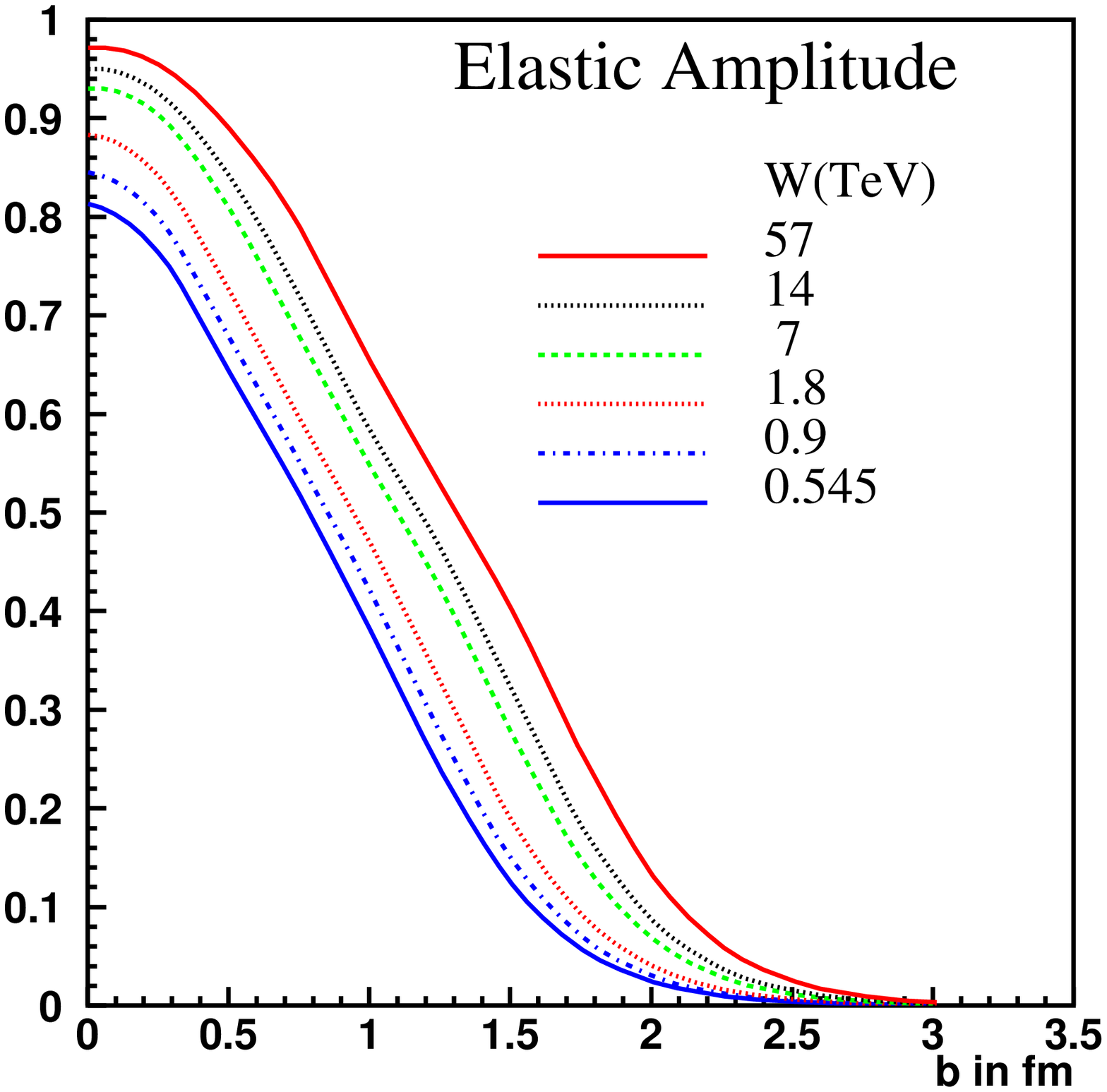}\\
\fig{amp}-c &  &\fig{amp}-d\\
\end{tabular}
\caption{ Amplitudes for one and two channel model  versus
 $b$ at different energies.  \fig{amp}-a shows the amplitude
 $A_{i k}(b)$ at W=7 TeV  as a function of the impact parameter
 $b$ in our model. The same amplitudes appear
 in \fig{amp}-b
 for the two channel model based on the Pomeron interaction\cite{GLMREV}.
 The energy behavior of the elastic amplitude
  is plotted in \fig{amp}-a and \fig{amp}-c 
 for the  one and two channels models, respectively.
 The figure for the one channel model is taken from Ref.\cite{GLMNIM}.}
\label{amp}
\end{figure}

We would like to draw the reader's attention to the energy behavior of the 
`black' 
$A_{1,1}$ amplitude, as  shown in \fig{a11w}. We note that the 
size of the black part of the amplitude for our model, is much
 smaller than for the Pomeron based model, discussed in  \cite{GLMREV}.
 The second observation is that this amplitude becomes transparent
 at low energies . One of the unpleasant features of the Pomeron (Regge 
pole) 
models
 is the fact that the black component already  appears at very low 
energies.
 The present model does not have this deficiency, providing a smooth
 transition from transparent to the black disc picture.

Comparing \fig{amp}-c and \fig{amp}-d we conclude that the 
two channel model leads to  weaker shadowing corrections than 
the one channel model.

Looking at \fig{amp}, it is difficult to avoid the pessimistic
 conclusion, that until we  know the non perturbative QCD
 structure of the hadrons, the high energy interaction of hadrons
 cannot be treated in a unique fashion, but  is doomed to be sensitive
 to {\it ad hoc}  assumptions.

\begin{figure}[h]
\centering
\begin{tabular}{l l l }
\includegraphics[width=0.4\textwidth,height=5.5cm]{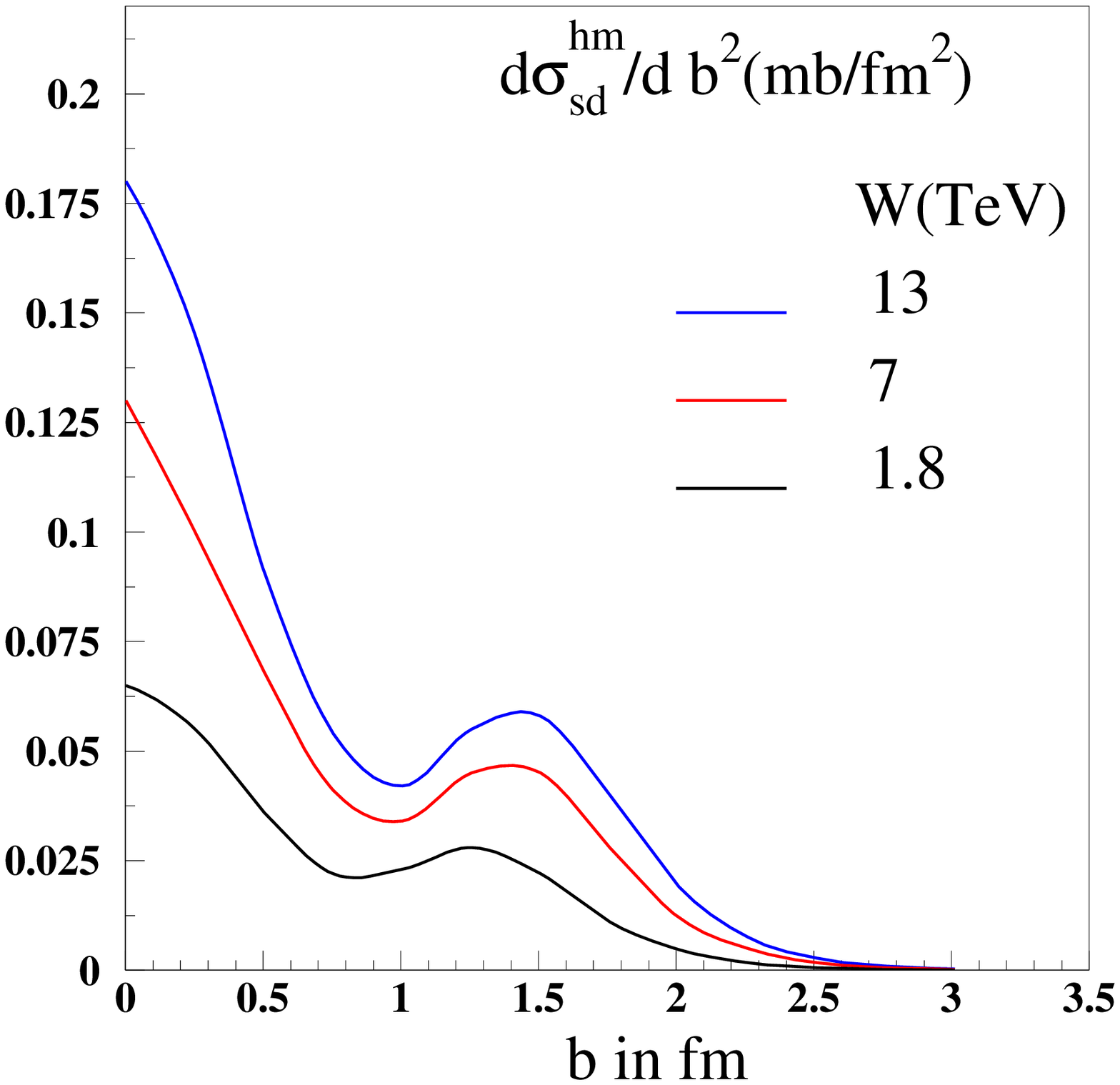}&~~~&\includegraphics[width=0.4\textwidth,height=5.5cm]{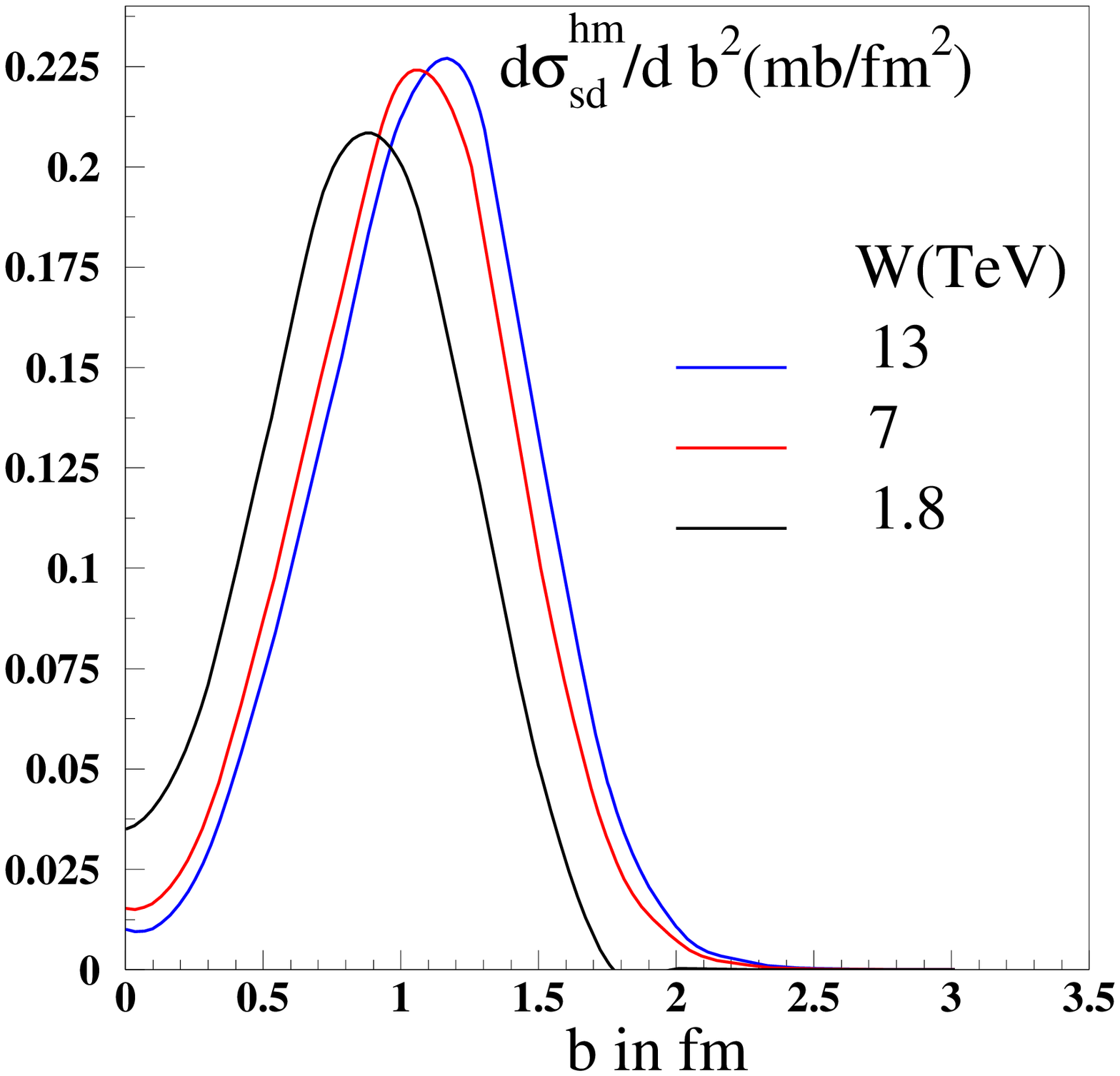}\\
\fig{damp}-a: high mass single diffraction  &   &\fig{damp}-b: low mass single diffraction\\
\includegraphics[width=0.4\textwidth,height=5.6cm]{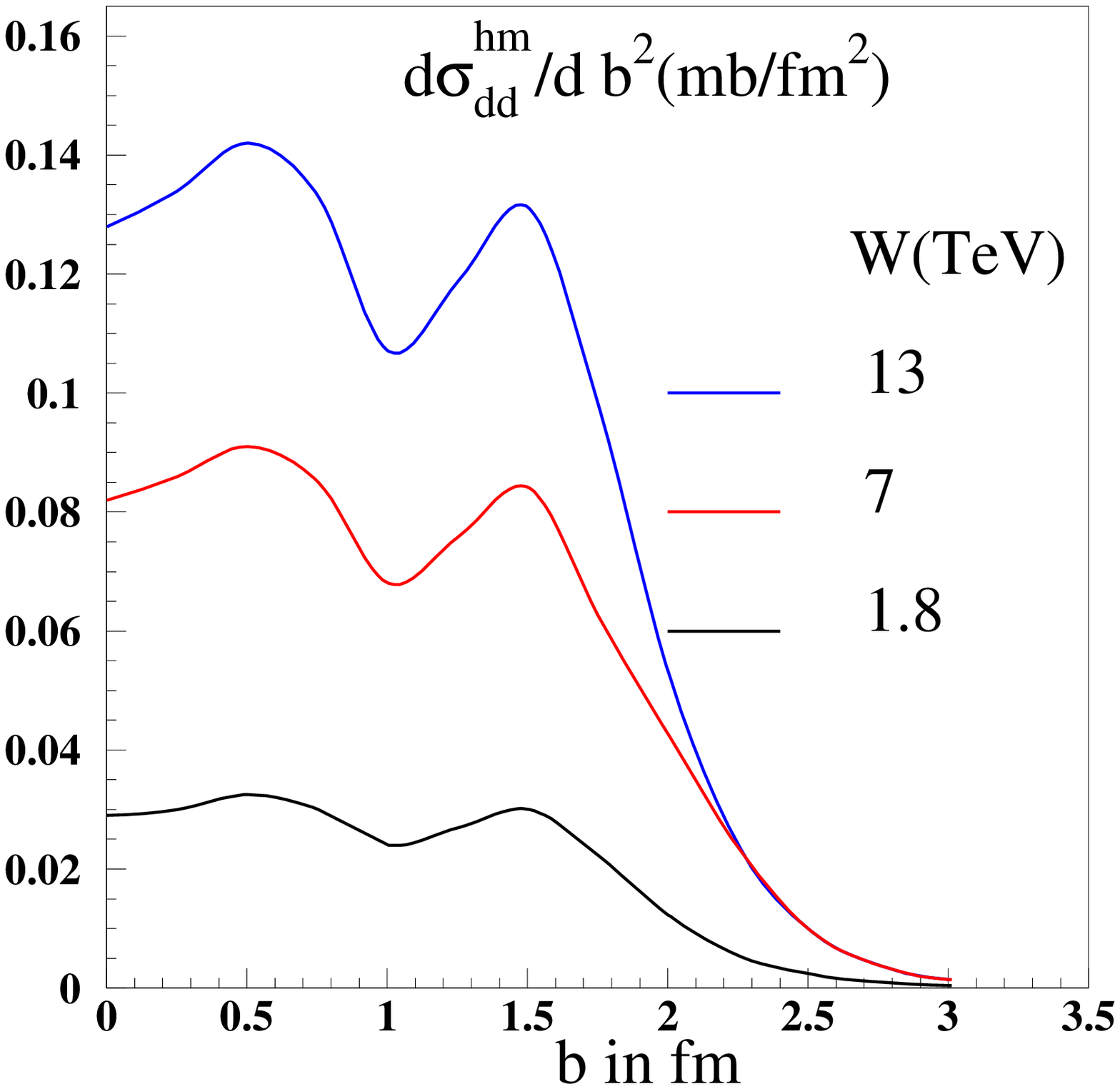}&  &
 \includegraphics[width=0.4\textwidth,height=5.5cm]{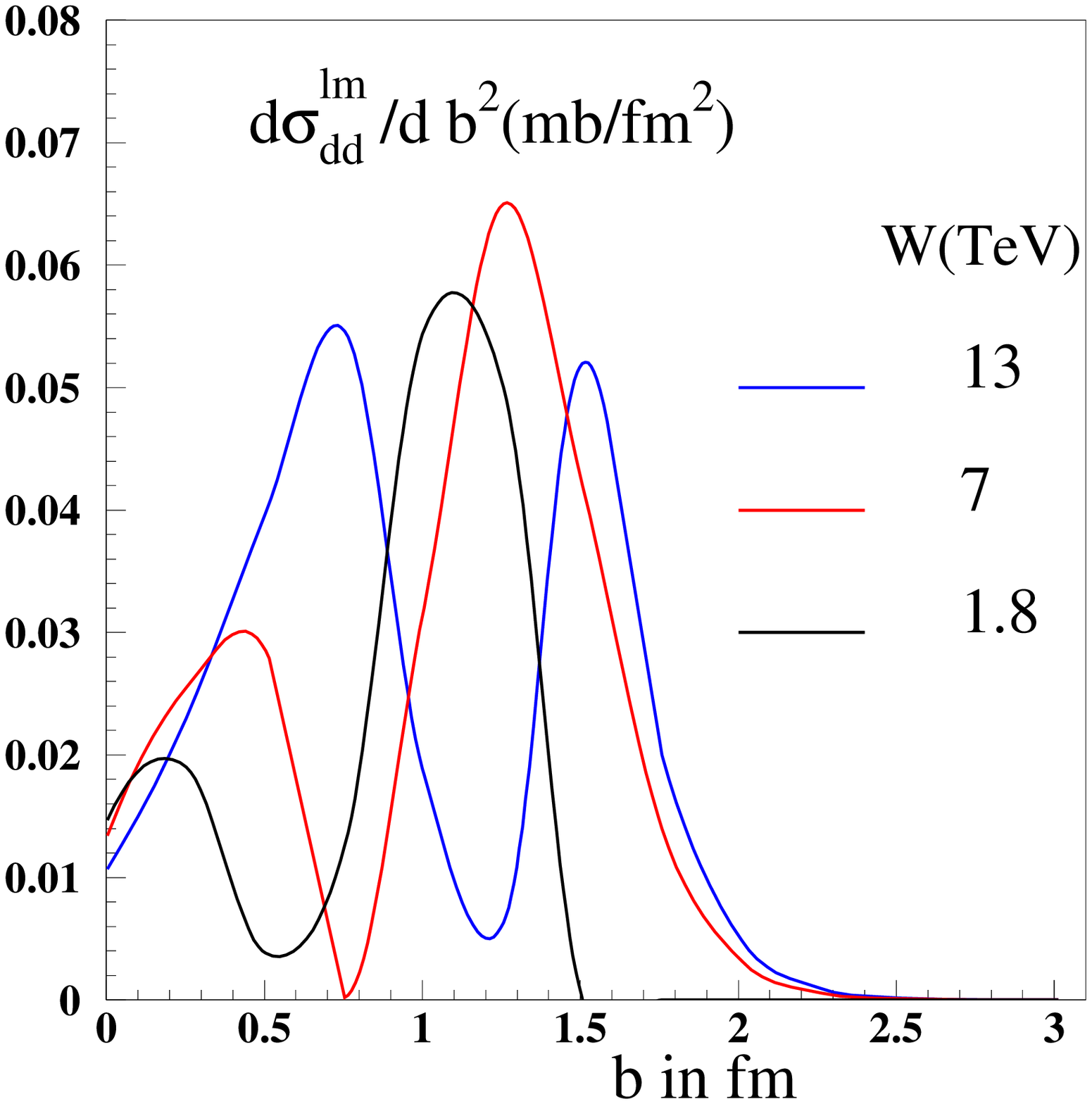}\\ 
\fig{damp}-c: high mass double  diffraction &  &\fig{damp}-d: low mass double  diffraction\\
\includegraphics[width=0.4\textwidth,height=5.6cm]{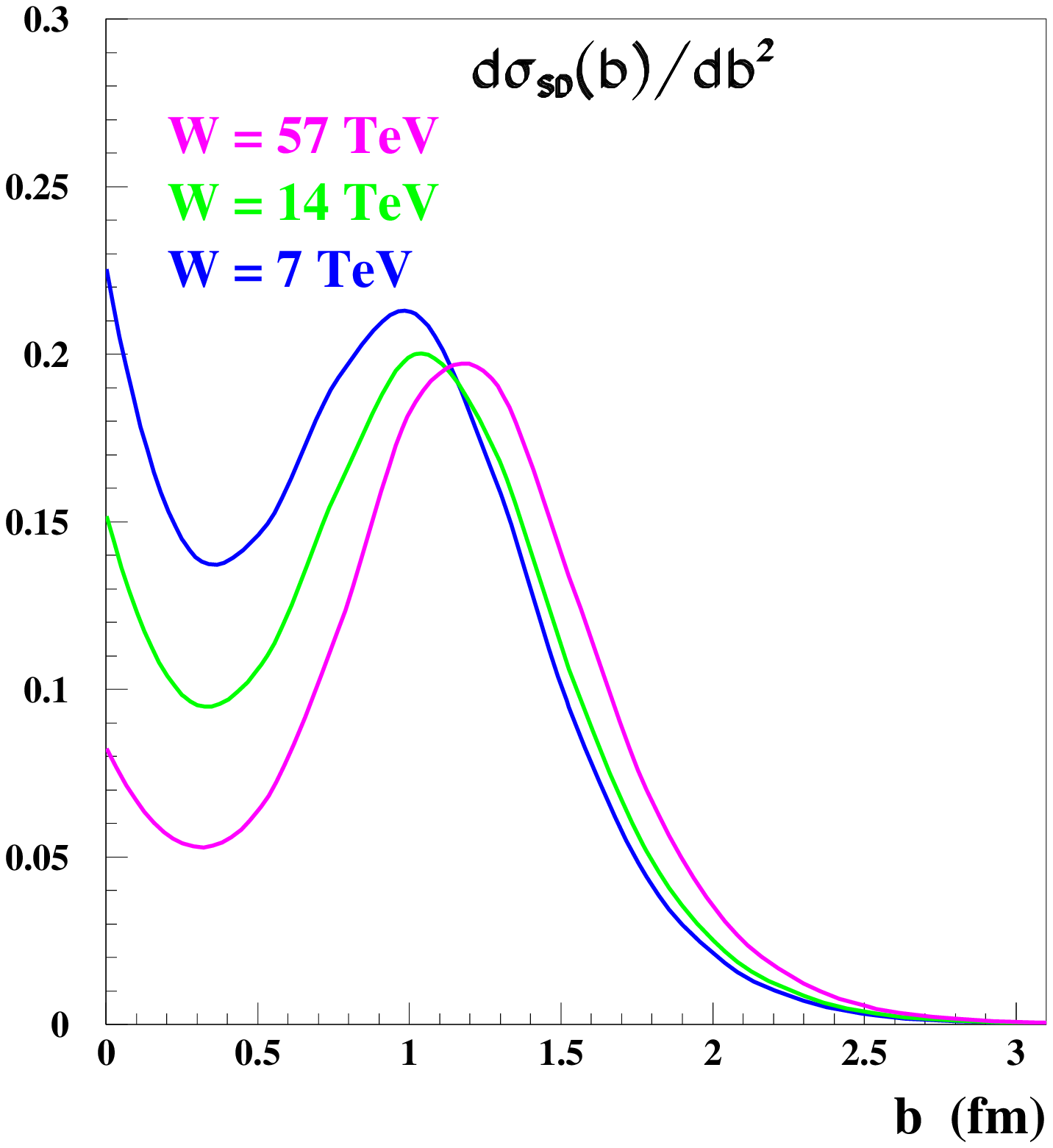}&  &
 \includegraphics[width=0.4\textwidth,height=5.5cm]{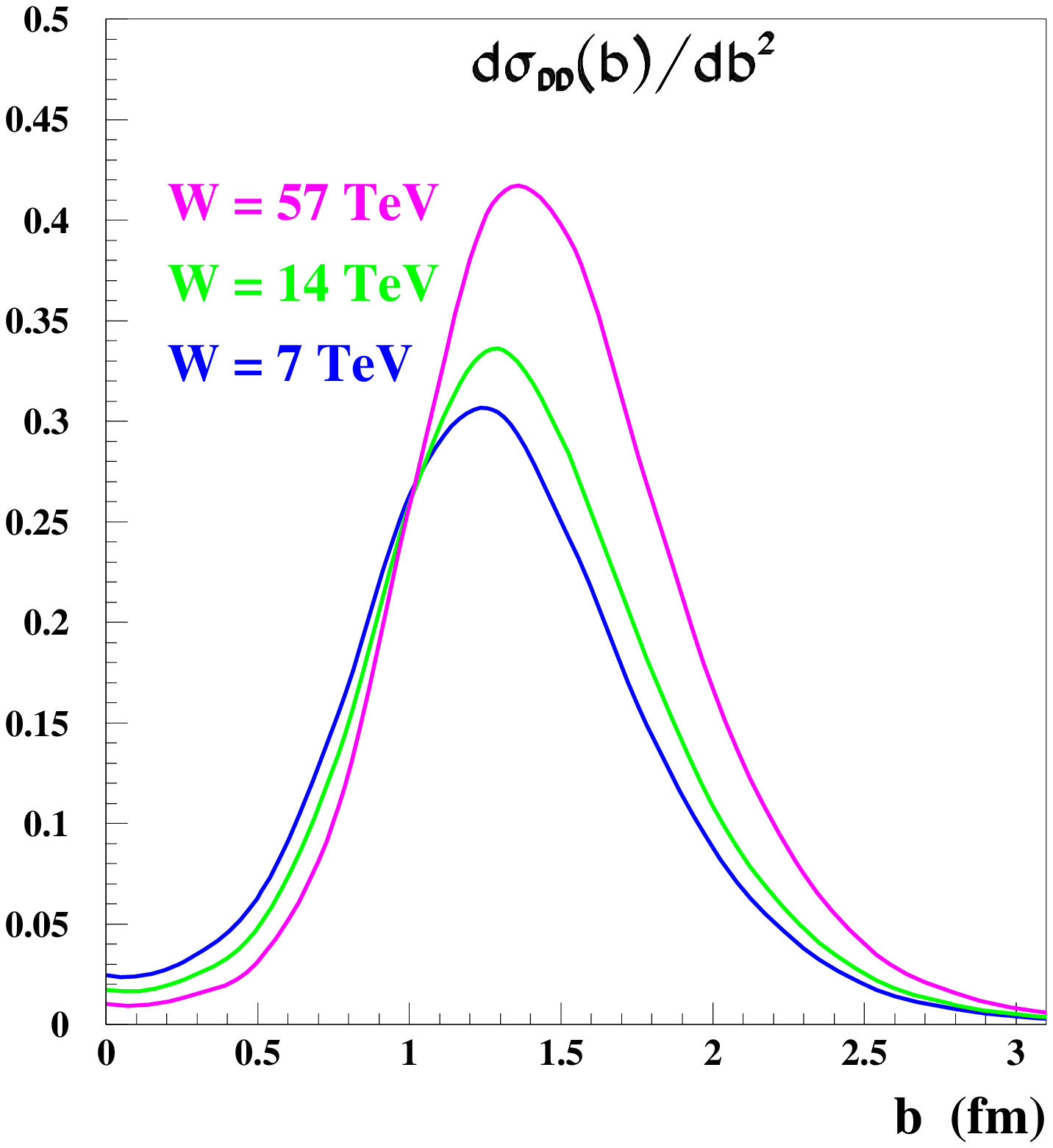}\\ 
\fig{damp}-e: single  diffraction (1 ch. model)  &  &\fig{damp}-f: double  diffraction (1ch. model)\\
\end{tabular}
\caption{ $d \sigma/d b^2$ versus b for different processes of diffraction production and different energies. `hm'(`lm') and `sd' (`dd') denote diffraction in high mass (low mass)  region  and single (double) diffraction respectively. \fig{damp}-e and \fig{damp}-f are taken from our paper in one channel model based on CGC/saturation approach\cite{GLMNIM}.
}
\label{damp}
\end{figure}

This statement can be illustrated by \fig{damp} where we plot the
 $d \sigma/d b^2$ for the  diffraction  production processes.   
  The low mass diffraction has pronounced peripheral features,
 having  minima at b=0 and at  b = 1 $\div$ 1.5 fm. However, for high
 mass diffraction the main contribution stems from b=0. Single 
diffraction
 in the region of high mass does not have a minimum at b=0, while in 
double
 diffraction such a minimum is seen, but it is very shallow.  Comparing 
these
 distributions with the one channel model, we see that for single
 diffraction, the pattern of the impact parameter distribution is 
similar
 in both cases. Double diffraction has a very clear peripheral
 distribution in the one channel model, but in the two channel model, the
 double diffraction dependence on  $b$ is similar to the single
diffraction. We believe that this comparison shows, that double   
 diffraction is sensitive to the production mechanism and, in
 particular, to the model of the hadron structure.
 
 We would like to point out the similarity of the distributions of our two 
channel dipole model (shown in Fig.7), with those obtained by the Durham 
group
\cite{NEWKMR}, inspite, of the difference in  the basic premises of the 
two groups.

\section{Conclusions}

In this paper we demonstrate that the generalization of the
 CGC/saturation based model to two channels, allows us to
 describe the experimental data, without having an oscillating
 behavior for the single diffraction production, and predicting
a value for the elastic slope at W=7\, TeV, which is smaller than the 
experimentally measured value.
  It should be stressed,
 that the values of the  phenomenological  parameters extracted from the
 fit satisfy the theoretical expectations. In addition we find  
new phenomenological small parameters: viz. $G_{3 \pom}/g_i \,\ll\,1$ and
 $m_i/m\,\ll\,1$ , which allows us to simplify the theoretical formulae. 

As  has been discussed above,  our approximation is valid only for
 energy less than (see \eq{MPSI2})
\beq \label{CON1}\
Y = \ln\Lb s/s_0\Rb \,\,\leq\,\,Y_{max} = \frac{2}{\Delta_{\mbox{\tiny BFKL}}}\,\ln\Big(\frac{1}
{\Delta^2_{\mbox{\tiny BFKL}}}\Big)\,\,\approx\,\,29
\eeq
for $\lambda =0.38$ (see Table I). Therefore, we can use our approach
 for energies up to $W = 100 \,TeV$.

We believe that this paper together with Ref.\cite{GLMNIM}, lends 
   support to the supposition, that a consistent model,  based on the 
BFKL
 Pomeron and the CGC/saturation approach, can be built. We have 
demonstrated
 that this model  successfully describes data for high energy
 hadron scattering.  In addition, we hope that this paper  provides  
credence to the 
arguments,
 that  the matching with long distance physics, (where the confinement of
 quarks and gluons is essential), can be reached within the CGC/saturation 
approach, this,
 without requiring that the soft Pomeron should appear (as a Regge pole).

  \section{Acknowledgements}
   We thank our colleagues at Tel Aviv university and UTFSM for
 encouraging discussions. Our special thanks go to Carlos Contreras,
 Alex Kovner and Misha Lublinsky for elucidating discussions on the
 subject of this paper.
   This research was supported by the BSF grant   2012124  and the
 Fondecyt (Chile) grant 1140842.

\end{document}